%% file: ATC_frac.tex
\documentclass[sigplan,10pt,urlbreakonhyphens=false]{acmart}
\usepackage{stfloats}
\usepackage{balance}
\AtBeginDocument{%
  }

\copyrightyear{2026}\acmYear{2026}
\setcopyright{cc}
\setcctype{by}
\acmConference[SIGOPS ATC 2026]{The 2026 ACM SIGOPS Annual Technical Conference}{November 15-18, 2026}{Hong Kong, China}
\acmBooktitle{The 2026 ACM SIGOPS Annual Technical Conference (SIGOPS ATC 2026), November 15-18, 2026, Hong Kong, China}
\acmDOI{10.1145/3828523.3857070}
\acmISBN{979-8-4007-2818-1/2026/11}

\acmSubmissionID{987}

\usepackage{xspace}
\usepackage{subcaption}
\usepackage[inline]{enumitem}
\usepackage{pifont}
\newcommand{\cmark}{\ding{51}}%
\newcommand{\xmark}{\ding{55}}%
\newcommand{\offrac}{f\textsc{RAC}\xspace}

\usepackage[normalem]{ulem}

\usepackage{scalerel}
\usepackage{tikz}
\usetikzlibrary{svg.path}

\input{KAUST_Logo}

\newif\ifsubmission
\submissiontrue

\ifsubmission
\newcommand{\mcnote}[1]{}
\newcommand{\sfnote}[1]{}
\newcommand{\zinote}[1]{}

\newcommand{\rsnote}[1]{}
\newcommand{\kinote}[1]{}

\newcommand{\removenote}[1]{}
\newcommand{\movetoappendix}[1]{}
\else
\usepackage{todonotes}
\newcommand{\mcnote}[1]{\todo[color=purple!40,inline]{MC: #1}}
\newcommand{\sfnote}[1]{\todo[color=blue!20,inline]{SF: #1}}
\newcommand{\zinote}[1]{\todo[color=yellow!40,inline]{ZI: #1}}

\newcommand{\rsnote}[1]{\todo[color=yellow!20,inline]{RS: #1}}
\newcommand{\kinote}[1]{\todo[color=gray!20,inline]{KI: #1}}

\newcommand{\removenote}[1]{\textcolor{red}{#1}}
\newcommand{\movetoappendix}[1]{\textcolor{red}{#1}}
\fi

\DeclareMathOperator{\Exp}{\text{Exp}}

\ifsubmission
 \usepackage{xpatch}
 \xapptocmd\normalsize{%
  \abovedisplayskip=0pt plus 1pt minus 1pt
  \abovedisplayshortskip=0pt plus 1pt
  \belowdisplayskip=0pt plus 1pt minus 1pt
  \belowdisplayshortskip=1pt plus 1pt minus 1pt
 }{}{}
\fi

\makeatletter
\renewcommand\acmBadgeR[2][]{%
  \ifx\@acmBadgeR\@empty
    \gdef\@acmBadgeR{%
      \smash{%
        \raisebox{-10pt}{%
          \href{#1}{\includegraphics[width=\@ACM@badge@width]{#2}}}}}%
  \else
    \g@addto@macro{\@acmBadgeR}{%
      \hspace{\@ACM@badge@skip}%
      \smash{%
        \raisebox{-10pt}{%
          \href{#1}{\includegraphics[width=\@ACM@badge@width]{#2}}}}}%
  \fi}
\makeatother
\acmBadgeR[https://www.acm.org/publications/policies/artifact-review-and-badging-current]{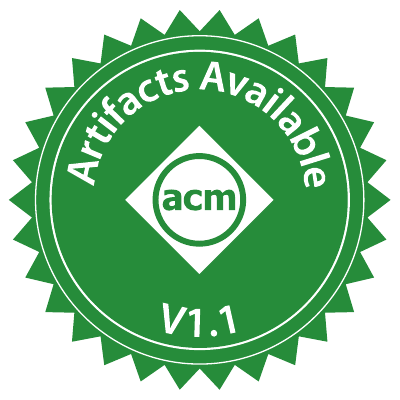}
\acmBadgeR[https://www.acm.org/publications/policies/artifact-review-and-badging-current]{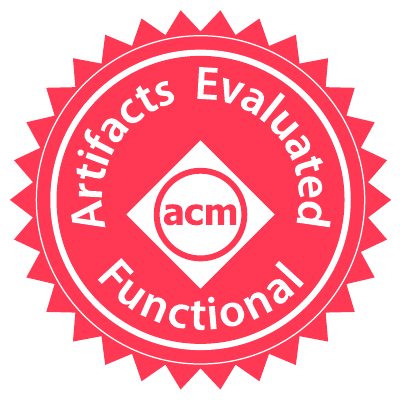}
\acmBadgeR[https://www.acm.org/publications/policies/artifact-review-and-badging-current]{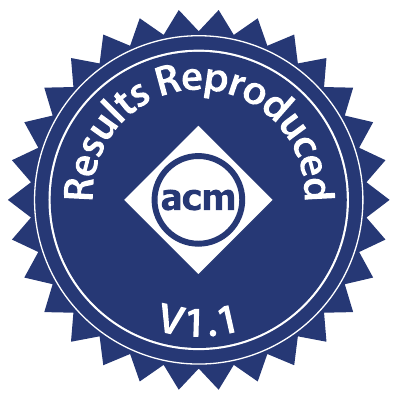}

\begin{document}
\pagestyle{plain}  % page number centered in the footer, no running heads
\raggedbottom

%%
%% The "title" command has an optional parameter,
%% allowing the author to define a "short title" to be used in page headers.
\title{fRAC: Remote Accelerator Calls on Hostless FPGAs}

%%
%% The "author" command and its associated commands are used to define
%% the authors and their affiliations.
%% Of note is the shared affiliation of the first two authors, and the
%% "authornote" and "authornotemark" commands
%% used to denote shared contribution to the research.
\author{Ziyi Yang}
\authornote{Both authors contributed equally to this work.}
\orcid{0000-0002-1372-9672}
\affiliation{%
  \institution{KAUST~\kausticon}
  \city{Thuwal}
  \country{Saudi Arabia}
}
\email{ziyi.yang@kaust.edu.sa}

\author{Krishnan B. Iyer}
\authornotemark[1]
\orcid{0000-0002-4608-329X}
\affiliation{%
  \institution{KAUST~\kausticon}
  \city{Thuwal}
  \country{Saudi Arabia}
}
\email{krishnan.iyer@kaust.edu.sa}

\author{Yixi Chen}
\orcid{0009-0003-0538-5667}
\affiliation{%
  \institution{KAUST~\kausticon}
  \city{Thuwal}
  \country{Saudi Arabia}
}
\email{yixi.chen@kaust.edu.sa}

\author{Ran Shu}
\authornote{Now at Broadcom.}
\orcid{0000-0002-2021-4917}
\affiliation{%
  \institution{Microsoft Research}
  \city{Beijing}
  \country{China}
}
% \email{ran.shu@microsoft.com}

\author{Zsolt Istv\'an}
\orcid{0000-0002-4127-8573}
\affiliation{%
  \institution{TU Darmstadt}
  \city{Darmstadt}
  \country{Germany}
}
\email{zsolt.istvan@tu-darmstadt.de}

\author{Marco Canini}
\orcid{0000-0002-5051-4283}
\affiliation{%
  \institution{KAUST~\kausticon}
  \city{Thuwal}
  \country{Saudi Arabia}
}
\email{marco@kaust.edu.sa}

\author{Suhaib A. Fahmy}
\orcid{0000-0003-0568-5048}
\affiliation{%
  \institution{KAUST~\kausticon}
  \city{Thuwal}
  \country{Saudi Arabia}
}
\email{suhaib.fahmy@kaust.edu.sa}

%%
%% By default, the full list of authors will be used in the page
%% headers. Often, this list is too long, and will overlap
%% other information printed in the page headers. This command allows
%% the author to define a more concise list
%% of authors' names for this purpose.
% \renewcommand{\shortauthors}{Anonymous Author(s)}

%%
%% The abstract is a short summary of the work to be presented in the
%% article.
\begin{abstract}
Hardware accelerators are essential for meeting microsecond-scale demands in distributed applications, but invoking them over the network remains bottlenecked by host software.
Host-mediated paths (RPC, RDMA) still rely on the CPU to reassemble requests and launch accelerators, which dominates end-to-end latency, while packet-level offload on SmartNICs lacks any notion of a complete request.
We introduce the Remote Accelerator Call (RAC) paradigm, in which transport ingestion, request reassembly, dispatch, and accelerator invocation are handled entirely in hardware.
We build \offrac, a hostless prototype that implements this network-to-accelerator interface as a lightweight shim on an FPGA hosting multiple virtualized accelerators, reconfigurable at runtime in a few milliseconds.
Serverless functions backed by fine-grained accelerators are a natural fit, since host-mediated invocation overhead can match or exceed their short execution times.
\offrac achieves under 7.8\,$\mu$s end-to-end latency while sustaining application throughput of up to 90\,Gbps.
Compared to highly optimized DPDK-based CPU and DPU baselines, it achieves at least 80\% lower latency under saturation, while consuming 85\% less power.
\end{abstract}


% %%
% %% Keywords. The author(s) should pick words that accurately describe
% %% the work being presented. Separate the keywords with commas.

% \received{20 February 2007}
% \received[revised]{12 March 2009}
% \received[accepted]{5 June 2009}

%%
%% This command processes the author and affiliation and title
%% information and builds the first part of the formatted document.
% \begingroup
% \mathchardef\UrlBreakPenalty=10000
\maketitle
% \endgroup

\section{Introduction}
\label{sec:intro}
Serverless platforms increasingly compose distributed applications from short-lived, compute-intensive functions.
While hardware accelerators (GPUs, FPGAs, ASICs) are well suited to executing these workloads with high efficiency~\cite{hennessy2019golden, jouppi2017datacenter, caulfield2016cloud,Dong2022Serverless}, their integration into distributed environments remains bottlenecked by software-mediated invocation paths.
Because serverless functions are inherently distributed, their inputs must be delivered over the network from clients or peer services.
When integrating accelerators to process these networked requests, current architectures adapt existing communication paradigms.
These include high-level RPC frameworks (standard in serverless platforms) and low-level RDMA (often used in high-performance accelerator clusters).
Across all these approaches, moving data to accelerators and invoking them requires software intervention.

\begin{figure}[t]
  \centering
  \includegraphics[width=0.95\columnwidth]{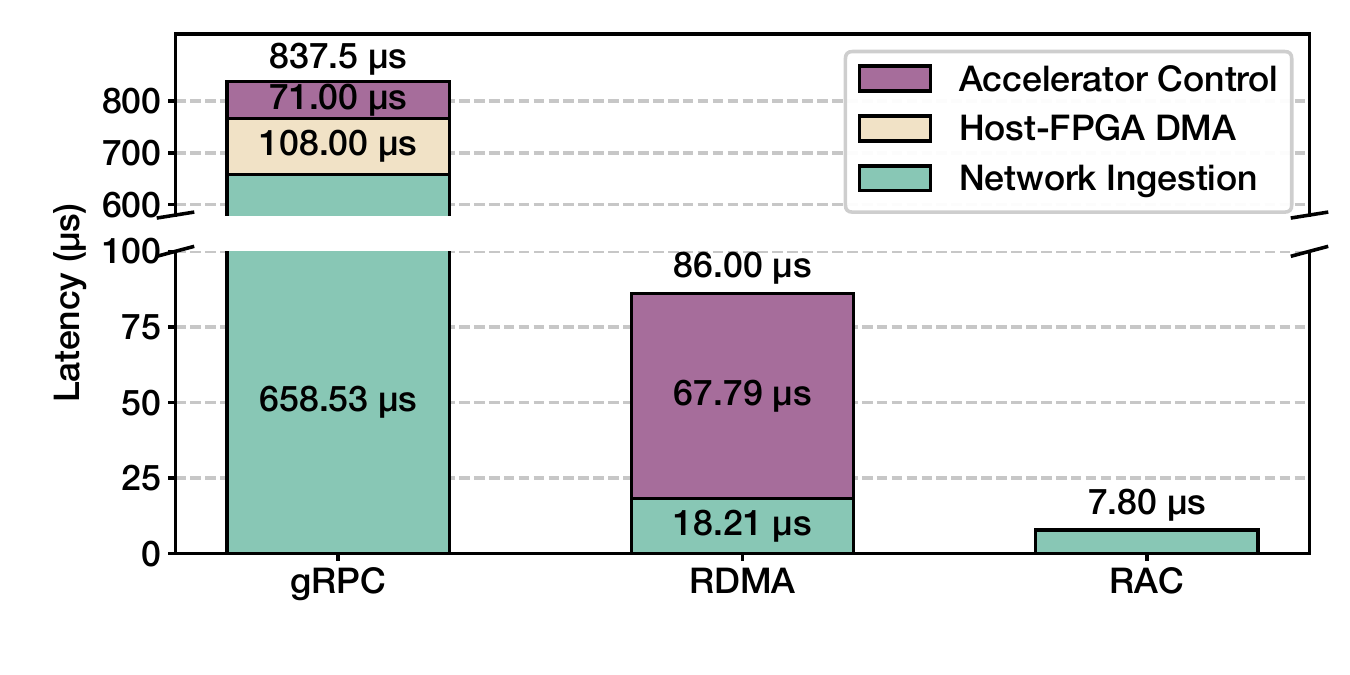}
  \caption{Latency breakdown of remote accelerator invocation for a 4KB payload across three paradigms. Host-Controlled (gRPC) is dominated by software network and RPC processing. RDMA with a host-side accelerator runtime optimizes the data path but leaves accelerator invocation under software control. RAC handles ingestion, dispatch, and invocation in hardware, eliminating the host-side segments altogether.}
  \label{fig:invocation_breakdown}
\end{figure}

In standard RPC frameworks, host software parses requests,
recovers request boundaries from fragmented packets, deserializes data, and schedules accelerator invocation~\cite{erpc,karandikar_hardware_2021,crankshaw2017clipper}.
Each of these steps consumes CPU cycles that scale with both payload size and request rate, and the resulting per-request software overhead is often comparable to or larger than the accelerator's computation time for microsecond-scale functions.
Even when data moves over PCIe to host-attached GPUs, FPGAs, and ASICs, software is still needed to configure DMA, detect transfer completion, and launch kernels through runtimes such as CUDA~\cite{cuda} or XRT~\cite{amdxrt}, each of which adds tens of microseconds of queue management and doorbell overhead per invocation \cite{Hou2024pcie_composable,mainas_f3_2025,Dong2022Serverless,cooke2020characterizing}.
To reduce these data-copy overheads, RDMA-capable NICs can DMA payloads directly into accelerator memory, whether GPU DRAM via GPUDirect RDMA or FPGA buffers exposed by an accelerator runtime~\cite{soNUMA,yuan_rambda_2023,amdxrt}.
However, RDMA exposes device or memory-region endpoints rather than request-granular accelerator endpoints.
It can place data into memory, but does not natively dispatch complete requests across multiple accelerator instances or trigger selected accelerators when requests are ready.
The host CPU still posts work requests, polls completion queues, and issues kernel launches.
Accelerator deployment frameworks additionally place request routing, scheduling, and device management on the software invocation path, adding overhead beyond data movement itself~\cite{Dong2022Serverless,mainas_f3_2025}.
Host mediation also keeps the CPU and its host platform active, adding substantial power overhead to otherwise lightweight accelerator invocations.
This host-managed model works well for the workloads that GPUs and TPUs are typically deployed for, such as large-batch, long-running computation (e.g., training and inference on large tensors), where the per-invocation control overhead is amortized over substantial compute time. For short-lived, microsecond-scale functions, however, this overhead remains on the critical path and becomes the dominant bottleneck once data movement is no longer the limit (Figure~\ref{fig:invocation_breakdown}).

Packet-processing hardware (SmartNICs, programmable NICs, and switch-based offload) keeps traffic in the datapath, but operates at packet granularity. Each packet is treated independently, with no notion of application-level requests, requiring accelerators themselves to manage payload buffering and invocation on a per-accelerator basis~\cite{ibanez2019p4netfpga,blocher2021switches}.
At the transport layer, offloads such as TCP Offload Engines (TOEs)~\cite{mogul2003tcp} can reconstruct reliable, ordered byte streams in hardware, but these streams do not expose application request boundaries or accelerator-invocation semantics.
Thus, across all these approaches, data movement may be optimized in hardware, but request formation and accelerator invocation remain on a software-run critical path.

A hardware path that connects transport ingestion directly to accelerator execution remains absent~\cite{fahmy2025hero}. Such a path must provide four capabilities without placing software on the critical path.
First, from the reliable, in-order byte stream provided by the transport layer, it must identify request boundaries and form complete application-level invocation requests.
Second, it must dispatch complete requests to one of multiple accelerators while supporting multiple clients concurrently.
Third, it must trigger accelerator execution when a request is ready, rather than waiting for software control.
Fourth, it must virtualize accelerators rather than binding each network endpoint to a fixed accelerator.
We call this paradigm Remote Accelerator Calls (RAC), where transport-to-accelerator invocation is handled directly in hardware. RAC is broadly applicable to accelerators that can be coupled with a transport endpoint on the same silicon.

We instantiate this paradigm in \offrac, a hardware prototype that implements these capabilities for network-attached FPGA accelerators.
\offrac implements accelerator virtualization in hardware, exposing diverse accelerators through a common request-level invocation abstraction.
% \kinote{what does single device mean here? it can be fixed set of accelerators}
Rather than being custom-built for a fixed accelerator, this abstraction lets flexible requests identify and invoke one of multiple accelerators through the same hardware interface.
In contrast, RPC- and RDMA-based frameworks may optimize data movement to an accelerator, but they do not provide a hardware abstraction for selecting among and invoking multiple diverse accelerators.
% \kinote{what does multi-accelerator virtualization mean here I think it should be multi-accelerator virtualization -> invocation abstraction}
% \kinote{ data movement and invocation -> data movement, invocation and virtualization of accelerators}
They rather require such invocation abstraction to be implemented in host software.
While our prototype implementation is built atop TCP, \offrac is designed to be transport layer agnostic. Our primary focus is data movement, invocation and virtualization of accelerators.
Motivated by the above, we make the following contributions:
% \kinote{ accelerator invocation through a request-level invocation interface while preserving isolation across requests.}
\begin{itemize}
    \item We define the Remote Accelerator Call (RAC) paradigm and implement the \offrac prototype that decouples network data transfer from accelerator invocation through a request-level invocation interface while preserving isolation across requests.    % The interface operates at request granularity and preserves isolation across requests.
    %\item We define the Remote Accelerator Call (RAC) paradigm and implement an \offrac prototype that decouples network transfer from accelerator invocation, operating at request granularity while preserving isolation.
    \item We show that \offrac adds only modest FPGA resource overheads while delivering high performance, achieving application throughput of up to 90~Gbps with end-to-end invocation latency under 7.8~$\mu$s, and enabling accelerator reconfiguration in under 10~ms.
    \item We evaluate \offrac against highly optimized DPDK-based CPU and DPU implementations, demonstrating at least 80\% lower latency than both platforms in high-throughput scenarios.
    \item We show that programmer effort to port existing accelerators to \offrac is minimal, as most designs already expose a suitable streaming interface.
\end{itemize}

\section{Background and Motivation}
\label{sec:background}

\begin{figure*}[t]
  \centering
  \includegraphics[width=\textwidth]{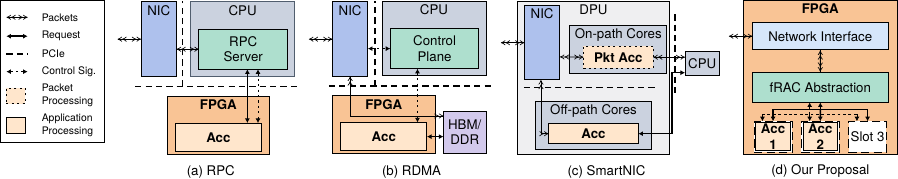}
  \caption{A comparison of network-to-accelerator invocation paths.
  \textbf{(a) RPC}: host software handles network processing, request parsing, scheduling, and accelerator invocation.
  \textbf{(b) RDMA}: the NIC can DMA data into registered memory regions reachable by the accelerator, but request dispatch and accelerator invocation remain host-managed.
  \textbf{(c) SmartNIC}: programmable datapaths either process traffic on-path at packet granularity (e.g., ClickNP~\cite{Li2016ClickNP}, PANIC~\cite{lin2020panic}) or support request-level processing off-path by traversing PCIe (e.g., FlexDriver~\cite{FlexDriver2022Eran}). Neither natively reassembles complete application requests for direct on-path accelerator invocation.
  \textbf{(d) \offrac}: a hostless FPGA path reassembles network fragments into complete serverless-function requests and invokes accelerators at application granularity.}
\label{fig:arch_comp}
\end{figure*}

Small, compute-intensive functions are increasingly offloaded to hardware accelerators to meet strict performance demands.
However, integrating these into distributed environments requires an efficient network-to-accelerator path.
Figure~\ref{fig:arch_comp} contrasts RAC with prior offloading paradigms.
Existing approaches rely on host-mediated control (e.g., RPC, RDMA) which add significant software latency, or use packet-level processing (e.g., SmartNICs) which cannot handle ap\-pli\-ca\-tion-level requests efficiently.
RAC eliminates the host CPU from the critical path by directly ingesting network data, reassembling complete requests, and triggering accelerators entirely in hardware. We now discuss the rationale for this design and defer to Section~\ref{sec:related_work} for a detailed comparison with related work.

\subsection{The Cost of Software-Mediated Invocation}
To understand why software remains a persistent bottleneck, we examine how the semantic gap between existing network abstractions and accelerator execution models manifests as a performance penalty. A complete accelerator invocation requires not only moving the request payload to the accelerator, but also explicitly triggering execution and polling for completion.

Figure~\ref{fig:invocation_breakdown} compares three invocation paradigms for a 4KB request and isolates where the bottleneck shifts as software is removed from the path.
\textbf{Host-Controlled (gRPC)} pays for everything in software, including network ingestion, host-FPGA DMA, and kernel launch (over AMD XRT~\cite{amdxrt} in this case). Because RPC abstractions are designed for general-purpose software services, they incur serialization and context-switching overheads that dwarf the execution time of microsecond-scale hardware functions.
\textbf{RDMA}\footnote{Our measured RDMA baseline uses a Mellanox NIC to DMA request data directly into XRT-allocated FPGA P2P buffers, while FPGA kernel invocation is still performed through XRT.} removes much of the software cost for networking and data movement.
However, one-sided RDMA exposes only low-level read, write, and atomic operations~\cite{burke2021prism}; richer behavior is possible only through elaborate verb chaining~\cite{reda2022redn}. The host CPU must therefore still intervene to synchronize and launch execution through an accelerator runtime such as XRT~\cite{amdxrt} or CUDA~\cite{cuda}. This mismatch between data movement and accelerator invocation leaves software overhead as the dominant remaining bottleneck. Today, it can be avoided only by hardwiring a fixed-function streaming accelerator into the network datapath, sacrificing the ability to share, select among, or replace accelerators.
\textbf{RAC} directly bridges this gap by providing an abstraction that is both expressive enough to support application-level invocation requests and lightweight enough to be implemented entirely in hardware, removing the software control path while simplifying accelerator deployment, as illustrated in Figure~\ref{fig:arch_comp}d.

\begin{figure}[t]
  \centering
  \includegraphics[width=0.95\columnwidth]{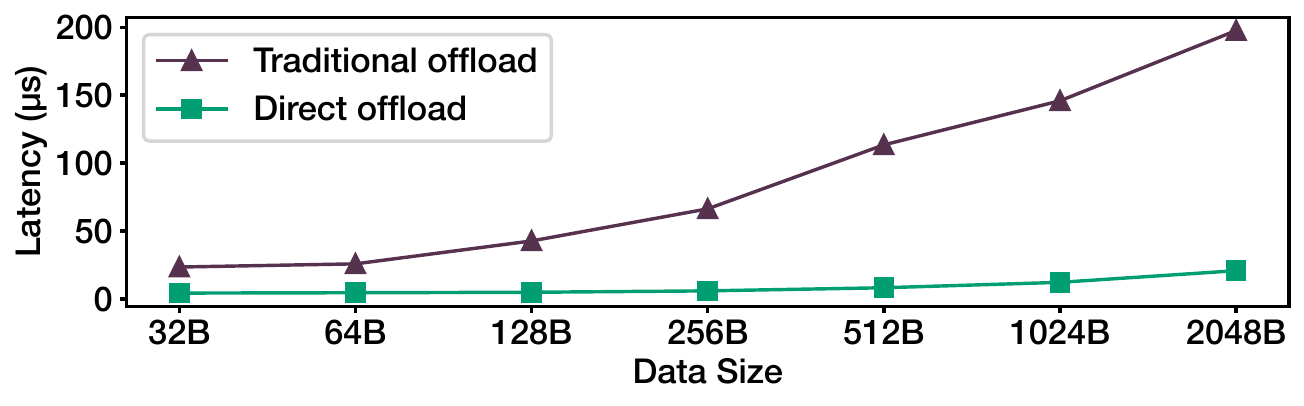}
  \caption{Median latency under host-controlled data movement and direct accelerator invocation, isolating the payload-size-dependent cost of the host PCIe path.}
  \label{click_figure_compare}
\end{figure}

Figure~\ref{click_figure_compare} provides a complementary view of host-mediated data movement, isolating the PCIe penalty as payload size scales.
Using a data sketch workload (Count-Min and Top-K) in ClickNP~\cite{Li2016ClickNP} on Microsoft Catapult~\cite{Catapult}, we compare traditional host-controlled data movement between two nodes implementing the two functions against direct accelerator chaining via the network.
Relying on the host CPU to move data across PCIe imposes a latency penalty that scales poorly with larger payloads.
Direct invocation bypasses the host PCIe path entirely, keeping data on the network or device and mitigating this scaling bottleneck.
This reveals a fundamental limitation: as long as a host CPU bridges the network and the accelerator, invocation will bottleneck performance, especially when distributing across nodes. To unlock hardware acceleration for microsecond-scale serverless functions, the invocation model must bypass the host entirely, allowing the network to deliver data and trigger accelerator execution directly in hardware.

\subsection{Fragmented Request Invocation}
Application-level requests often have a granularity that exceeds the network MTU, leading to data being fragmented across multiple network packets.
While transport protocols transparently segment and reassemble an in-order byte stream, they do not preserve message boundaries. Hardware that operates directly on such a stream therefore faces a semantic mismatch between network-level packets and application-level requests.

Hardware accelerators typically cannot context switch mid-request, since their intermediate state is distributed across pipeline registers and on-chip buffers that cannot be cheaply saved and restored. Once an accelerator begins consuming a request, it must remain dedicated to that request until it completes. If request fragments are forwarded directly as they arrive from the network, interleaved requests cause head-of-line blocking: fragments of other requests are ready, but the accelerator cannot accept them. For example, a streaming image filter that has begun processing one image stalls whenever a subsequent packet is delayed by jitter or congestion, creating ``invocation bubbles''. In a multi-tenant setting, a stalled accelerator is a locked resource: it cannot serve other clients while it waits for a delayed packet. Prior RPC-offload designs reassemble requests before dispatch but still involve the host: NEBULA~\cite{sutherland_nebula_2020} delivers them to CPU-run handlers, and RpcNIC~\cite{zhang_rpcnic_nodate} runs reconfigurable kernels on a PCIe-attached SmartNIC under host-software control.

Eliminating these bubbles requires decoupling network delivery from accelerator execution. Incoming fragments should be buffered, and requests dispatched to accelerators only once they have been fully reassembled.
This absorbs network delay before invocation: once started, the accelerator streams the request without interruption and can immediately proceed to the next queued request.
Although each accelerator could buffer its own inputs, it would need a separate buffer for every concurrently arriving request, provisioned for the worst case and duplicated across every accelerator. Our key insight is to  lift reassembly out of individual accelerators into a shared, accelerator-agnostic layer, where buffer capacity is multiplexed across tenants and any streaming accelerator can be deployed without modification.

\begin{table*}[t!]
  \caption{Examples of accelerators used in distributed applications and their FPGA implementations.}
  \label{tab:acc-functions}
  \small
  \centering
  \begin{tabular*}{\textwidth}{@{\extracolsep{\fill}}lllll@{}}
  \toprule
  \textbf{Category} & \textbf{Example Accelerators}  &\textbf{Data Granularity} &\textbf{Runtime} & \textbf{FPGA Impl.}\\ \midrule
  Audio/Video & Filtering, Resizing, Speech-to-text, Video encoding  & Frame ($\approx$ 200--500KB) & $\sim$2ms & \cite{Liu2020VoiceControl, Han2017LSTM, wan2021Resizing}\\
  Mathematical & Matrix Multiplication, SVD, Cholesky decomp., EMA & Matrix ($\approx$ 2--32KB)  &  $\sim$10\textmu s & \cite{de_Fine_Licht_2020_MM, Muno2015QR, Ma2006SVD, Luo2013Cholesky, Shome2012filter}\\
  Machine Learning & CNN inf., Clustering, Bayes classific., Aggregation & Tensor Block ($\approx$ 3KB--3MB) & $\sim$5ms& \cite{Jokic2018BinaryEye, bacis2020blastfunction, Pu2015KNN, Chou2020Bayes} \\
  Data Analysis & Top-K selection, Count-Min Sketch & Stream data & $\sim$1\textmu s/KB& \cite{Kulkarni2020HyperLogLog, Istvan2014Histograms, Matsumopto2015TopK}\\
  ML Pre-Processing & Logit Transform, Min-Max Normalization&  Tensor Block ($\approx$ 3KB--3MB) &  $\sim$3--30\textmu s & \cite{Hua2018Logit, Mohsin2018KNN} \\
\bottomrule
  \end{tabular*}
\end{table*}

\subsection{Serverless Functions with RAC}
Serverless computing is a natural fit for the RAC model: applications are decomposed into short-lived, highly concurrent functions invoked on demand. Crucially, each invocation operates on a complete input, such as an image or a tensor block, rather than on individual network packets. As Table~\ref{tab:acc-functions} shows, many such functions have efficient hardware implementations, and once accelerated, many complete in microseconds or milliseconds. For the shortest of these, host-mediated invocation can add latency comparable to the function itself, and for all of them, stalls caused by fragmented network delivery leave accelerators underutilized. These functions therefore require an execution model in which invocation granularity matches the input size and execution starts as soon as a complete input is available, which is precisely what RAC's complete-request invocation provides.

A serverless deployment separates orchestration from execution: a control plane places functions and manages workers, while workers serve invocation requests.
RAC allows these workers to be hostless accelerator nodes: requests are reassembled and dispatched directly to concurrent accelerator pipelines, without host involvement, serving multiple clients in isolation and without invocation bubbles.

\section{\offrac Design}\label{sec:design}
% \kinote{function invocation -> accelerator invocation}
\offrac is a lightweight framework that realizes RAC on net\-work-attached FPGAs. FPGAs are a practical substrate for RAC: a single device exposes multiple accelerator slots, and dynamic partial reconfiguration allows accelerators to be swapped at runtime to adapt to changing demand~\cite{amddfx,vipin2018fpga}. While most FPGA virtualization frameworks rely on host software to multiplex clients, manage execution, and reconfigure accelerators~\cite{chen2014enabling,fahmy2015virtualized,korolija2020OS}, \offrac performs these tasks on the FPGA itself.

\begin{figure*}[t]
\centering
\includegraphics[width=1.0\linewidth]{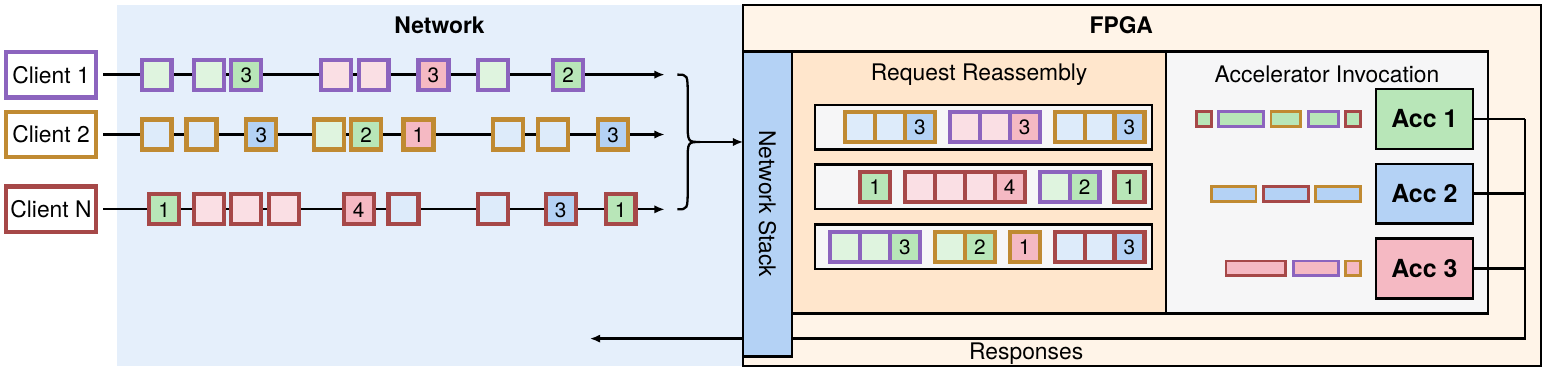}
\caption{Overview of \offrac operating model. Multiple clients (outline color) issue requests addressing different accelerators  (fill color), which are fragmented over multiple packets, the first of which determines the size.
These fragments are reassembled into complete requests and dispatched to the corresponding accelerator.
Reassembly is performed in buffers that are agnostic to client and accelerator for resource efficiency.
}
\label{fig:overview}
\end{figure*}

Incoming network data is reassembled into complete ap\-pli\-ca\-tion-level requests and buffered on-chip; each request is then routed directly to its target accelerator, which processes it to completion before serving the next. A reconfiguration controller swaps accelerators into slots on request from the control plane, keeping the network path active so that functions in other slots continue executing. Because \offrac is built atop a standard network stack (Figure~\ref{fig:overview}), it requires no changes to existing infrastructure.

In a serverless deployment, \offrac operates as a hostless worker behind an existing control plane and gateway, which handle orchestration, request routing, and platform security; we therefore focus on the worker's datapath. \offrac hosts a library of hardware functions rather than arbitrary FaaS code. Supported functions expose a streaming interface, keep their per-invocation working set on-chip, and fit within the resources of a single accelerator slot; persistent state remains outside the FPGA.

\offrac provides request-level data isolation: each reassembly buffer holds only one in-flight request, and accelerators run each request to completion, so data from different requests is never interleaved. It also isolates stalls: only complete requests are dispatched, so a slow sender holds a buffer rather than an accelerator, and a long-running function delays only its own queue.

The remainder of this section addresses three design challenges: buffer management, accelerator scheduling, and accelerator reconfiguration. We validate each design decision in simulation\footnote{The simulator setup is detailed in Appendix~\ref{apx:simulator} (Table~\ref{tab:simulator_setup}).} before presenting the implemented system in Section~\ref{sec:prototype}.

\subsection{On-Chip Request Reassembly}
\label{computation_model}

% Larger application requests are fragmented into multiple packets due to MTU constraints.
% Even assuming in-order delivery from the transport layer, there is no guarantee that these packets will arrive back-to-back.
% An accelerator cannot serve another request until it has processed all fragments of a current request, and hence is likely to remain idle for longer periods as request size
% %\removenote{(and, hence, fragmentation)}
% increases.
% This underutilization of resources significantly reduces the potential throughput benefits of the accelerator.

% As established in Section~\ref{sec:background}, feeding network fragments directly into an accelerator causes severe underutilization due to invocation bubbles.
% To address this, \offrac decouples network delivery from execution by reassembling requests on-chip based on a custom request header (inserted by the client) that indicates the request size (cf. Figure~\ref{fig:Request-format}).

% \offrac appends incoming fragments to a dedicated hardware reassembly buffer in on-chip FPGA memory (rather than external DRAM/HBM) to minimize latency. Once a request is entirely received, it is passed to the accelerator queues. Although avoiding external memory limits request sizes to single-digit MBs, the representative applications discussed in Section~\ref{sec:background} fit comfortably within this range.

To eliminate the invocation bubbles described in Section~\ref{sec:background}, \offrac reassembles requests on-chip before invoking any accelerator.
Reconstructing the byte stream remains the job of the transport layer; \offrac instead recovers request boundaries within that stream, in hardware, across many concurrent connections, so that a complete request becomes the unit of accelerator invocation.
The first fragment of each request carries a custom header specifying its total size and target function (cf. Figure~\ref{fig:Request-format}). On arrival \offrac allocates a reassembly buffer for that request from a shared pool, using on-chip FPGA memory rather than external DRAM/HBM to minimize latency, and appends fragments to it. Once the request is complete, it is forwarded to the target accelerator's queue.
Message framing itself is standard in software RPC frameworks; what \offrac contributes is performing it in the datapath of a hostless device and binding it directly to accelerator dispatch.

\begin{figure}[t]
  \centering
  \includegraphics[width=\columnwidth]{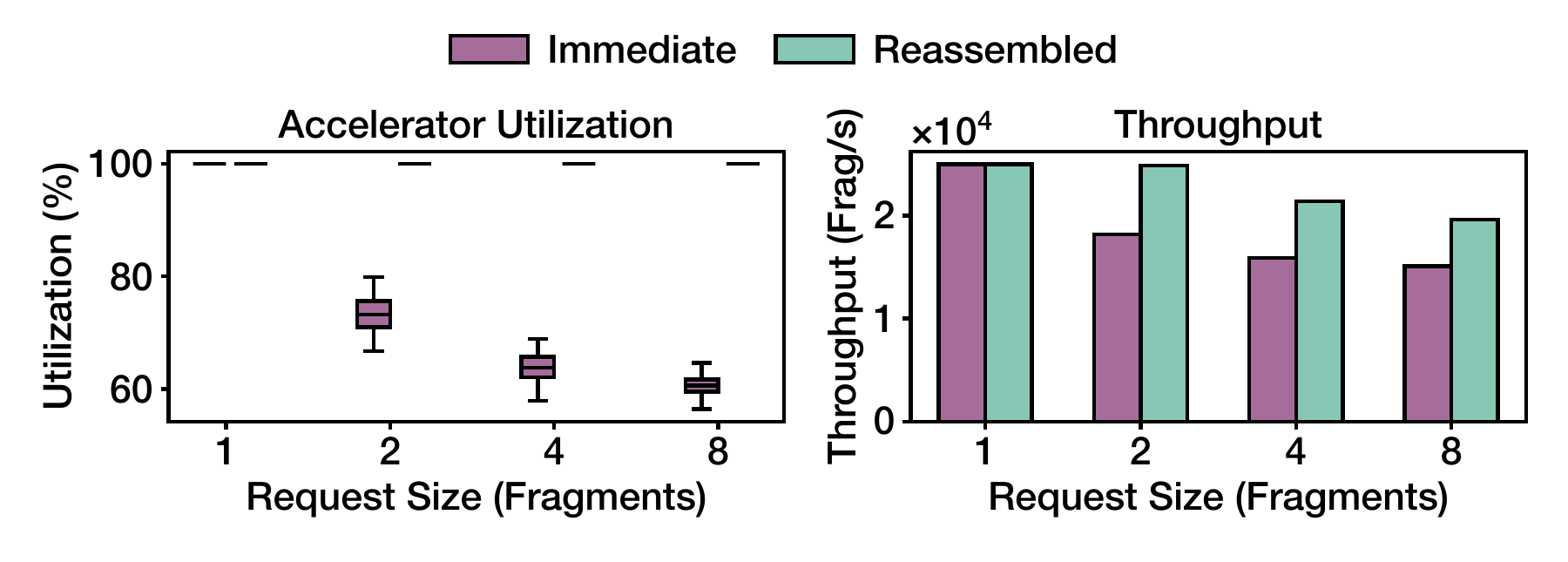}
  \caption{Comparison of immediate data ingestion at an accelerator vs. ingestion of reassembled requests.}
  \label{fig:reassembly-sim}
\end{figure}

We validate the impact of request reassembly on accelerator utilization and multi-client support with two simulations.
We align our setup with the function characteristics in Table~\ref{tab:acc-functions}, modeling an accelerator that processes requests composed of 1, 2, 4, or 8 fragments, with each fragment requiring 20 $\mu$s to process.
Fragments belonging to the same request arrive with a random inter-fragment delay of 30--40 $\mu$s.
The 20~$\mu$s processing time is representative of Table~\ref{tab:acc-functions}, and the slightly longer inter-fragment delay exposes invocation bubbles.
Figure~\ref{fig:reassembly-sim} (left) shows that the utilization of the accelerator, i.e., the time spent processing requests rather than stalling, decreases as request size increases when fragments are fed straight into the accelerator (Immediate in the plot). In contrast, enabling request reassembly (Reassembled) significantly improves utilization by eliminating stalling, since the accelerator only begins execution after receiving a complete request.

Eliminating stalling has the added benefit that an accelerator can service more requests from distinct clients.
We validate this with an experiment with two clients.
Without request reassembly, any new request arriving from one client while the accelerator is busy serving a request from the other is dropped.
Figure~\ref{fig:reassembly-sim} (right) shows that as request size increases, immediate ingestion sees decreasing throughput as the accelerator spends 25--40\% of its time stalled, while the design with reassembly can service queued complete requests back-to-back without stalling. Further analysis is in Appendix~\ref{apx:queue}.

Hence, by presenting accelerators with only complete requests using reassembly buffers, \offrac improves both accelerator utilization and request throughput in multi-tenant settings.

\subsection{Designing an Efficient Reassembly Buffer}\label{sec:bufferdesign}
We first ask \emph{how should reassembly buffers be coupled with accelerators?}
An intuitive approach is to adopt a per-accelerator reassembly buffer similar to the per-core shallow queue in~\cite{lin2023ringleader}.
However, this approach can lead to Head-of-Line (HoL) blocking when there are multiple clients issuing requests to the same accelerator.
A single reassembly buffer can only reassemble one request at a time to maintain the contiguous order of fragments that constitute requests.
This significantly limits scalability and leads to under-utilization of accelerators.
Furthermore, it would be necessary to allocate suitable buffer space per accelerator based on the properties of the accelerator (e.g., request size distribution and execution time) and expected demand.

Rather than statically allocating buffers to specific accelerators, \offrac decouples them, enabling reassembly to be optimized independently of accelerator and request size mix, as is necessary for a virtualized system where the specific accelerator loaded in a slot might differ over time.
As depicted in Figure~\ref{fig:overview}, each accelerator has a separate queue that complete requests are forwarded to after reassembly, and these are processed atomically as soon as the accelerator becomes available.

\offrac instantiates multiple reassembly buffers that operate as follows.
The first fragment of a new incoming request is dispatched to eligible buffers based on a policy, explored in Section~\ref{sec:simreqbuffalloc}. 
A buffer is \emph{eligible} to accept the first fragment of a new request only if it is not currently reassembling a request, and only if the size of the new incoming request is no larger than the remaining space in the buffer.
Once the first fragment of a request has been admitted to a buffer, that buffer is now allocated solely to the remaining fragments of that request; it is \emph{ineligible} to serve any other request.
Once a buffer completes assembling a request, it becomes eligible to accept a new request once more.
If there are no eligible buffers, the incoming request fails (as with the request's remaining fragments).
We omit the transport stack in our simulation to isolate the \offrac abstraction. In a full deployment, these ``failed'' fragments are absorbed by the reliable transport layer (e.g., TCP) and seamlessly trigger network retransmissions, rather than causing application-level failures.
Assembled requests are held in the buffer only until they can be sent to accelerator queues.

\begin{figure}[t]
  \centering
  \begin{subfigure}[b]{0.3\columnwidth}
      \centering
      \includegraphics[width=\textwidth]{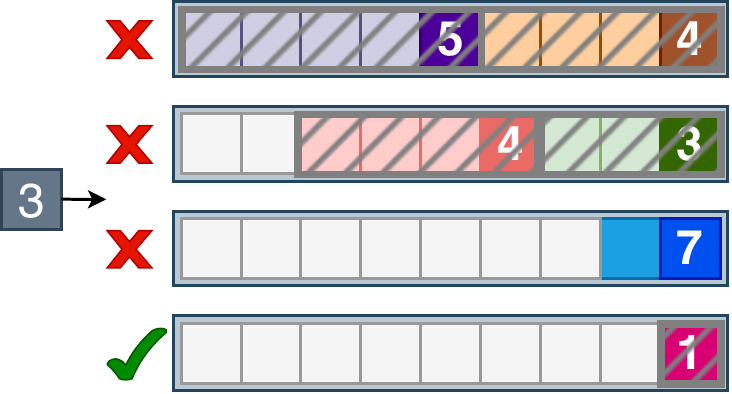}
      \caption{New request.}
      \label{fig:reassembly-policy-starting}
  \end{subfigure}
  \hspace{0.02\columnwidth}
  \begin{subfigure}[b]{0.3\columnwidth}
      \centering
      \includegraphics[width=\textwidth]{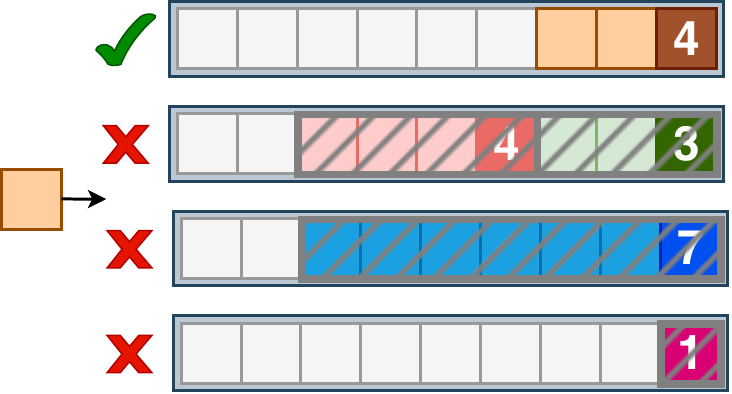}
      \caption{In-flight.}
      \label{fig:reassembly-policy-concatenation}
  \end{subfigure}
  \hspace{0.02\columnwidth}
  \begin{subfigure}[b]{0.3\columnwidth}
      \centering
      \includegraphics[width=\textwidth]{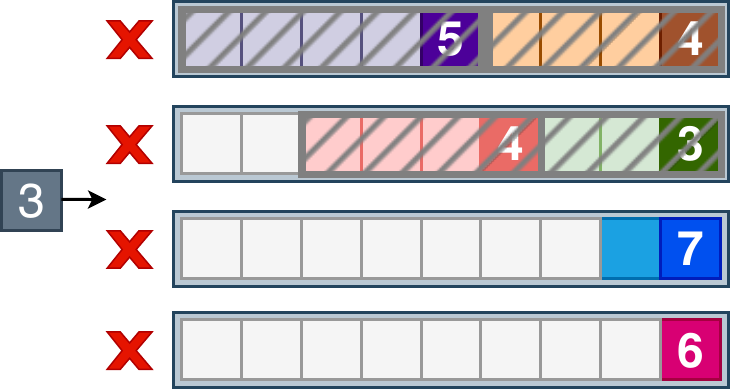}
      \caption{Failed request.}
      \label{fig:reassembly-policy-dropping}
  \end{subfigure}
  \caption{Behavior of the Reassembly Buffers.
  (a) The first fragment of each request identifies its size, and can be accepted only into an eligible buffer that is not reassembling another request and has sufficient capacity.
  (b) Subsequent fragments of a request are steered to the corresponding buffer.
  (c) If a new request does not find an eligible buffer, it fails.
  \emph{Legend: Fully assembled requests are hatched. Eligible buffers in each scenario are marked with a tick and ineligible buffers with a cross.}}
  \label{fig:reassembly-policy}
\end{figure}

Figure~\ref{fig:reassembly-policy} shows \offrac's behavior in response to buffer states and incoming fragments.
These policies are simple enough to be enforced at the hardware level using per-clock status signals.
However, \offrac relies on the underlying transport protocol to guarantee in-order data delivery.
If a connection is interrupted before a request is fully reassembled, \offrac must perform garbage collection to discard incomplete fragments from the reassembly buffer.

\begin{figure}[t]
    \centering
        \centering
        \includegraphics[width=\linewidth]{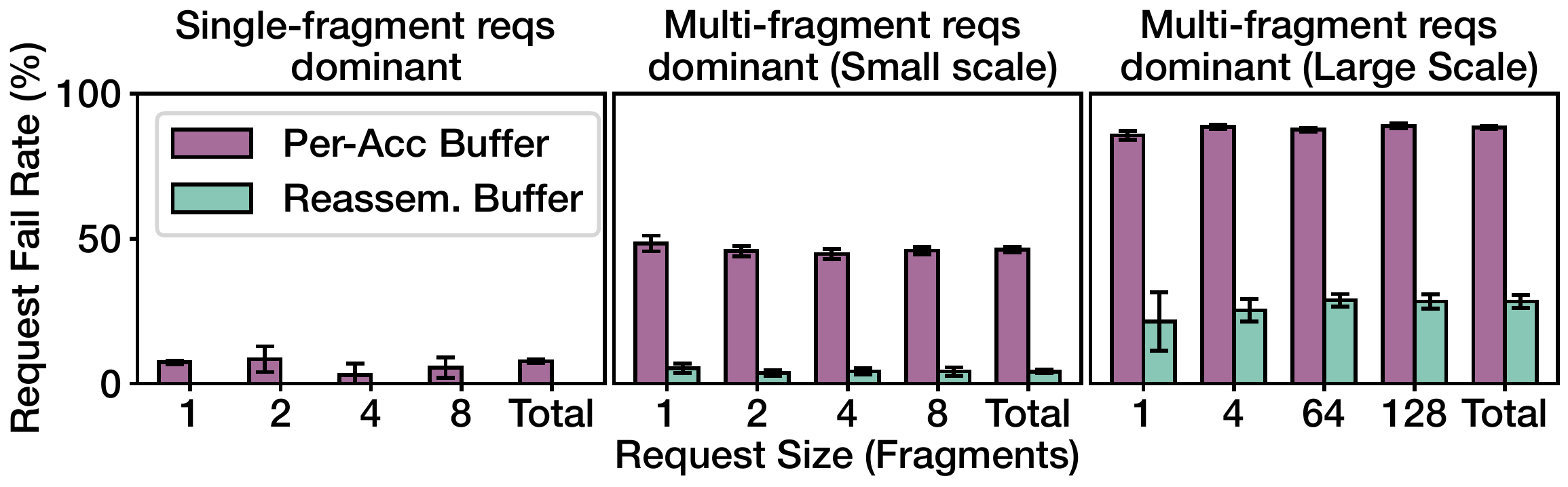}
        \caption{Request fail rates with per-accelerator buffers vs.\ a separate reassembly buffer.}
        \label{fig:separate-reassembly-buffer}
\end{figure}

We simulate per-accelerator and independent reassembly buffers when servicing diverse requests in a setup with four buffers and three accelerators.
Incoming requests are allocated in a round robin manner to \textit{eligible} buffers.
Three different workload mixes are used to explore the impact on request fail rate under a high saturating load: single-fragment requests dominate, and multi-fragment requests dominate (evaluated at both small and large request scales).
Figure~\ref{fig:separate-reassembly-buffer} shows that decoupling reassembly buffers from accelerators significantly reduces the request fail rate, dropping by approximately 40\% and 60\% for small-scale and large-scale multi-fragment workloads, respectively.
% Three different workload mixes are used: single-fragment requests dominate (i.e., reassembly is unnecessary), request sizes are evenly distributed, and multi-fragment requests dominate.
% We generate requests at a high saturating rate to explore the impact on request fail rate.
% Figure~\ref{fig:separate-reassembly-buffer} shows that decoupling reassembly buffers from accelerators significantly reduces request fail rate, by 42\% and 60\% respectively, for small-scale and large-scale multi-fragment requests.


\subsection{Dealing with Small Requests}\label{sec:simsinglereqbuff}
\begin{figure}[t]
  \centering
  \includegraphics[width=1.0\linewidth]{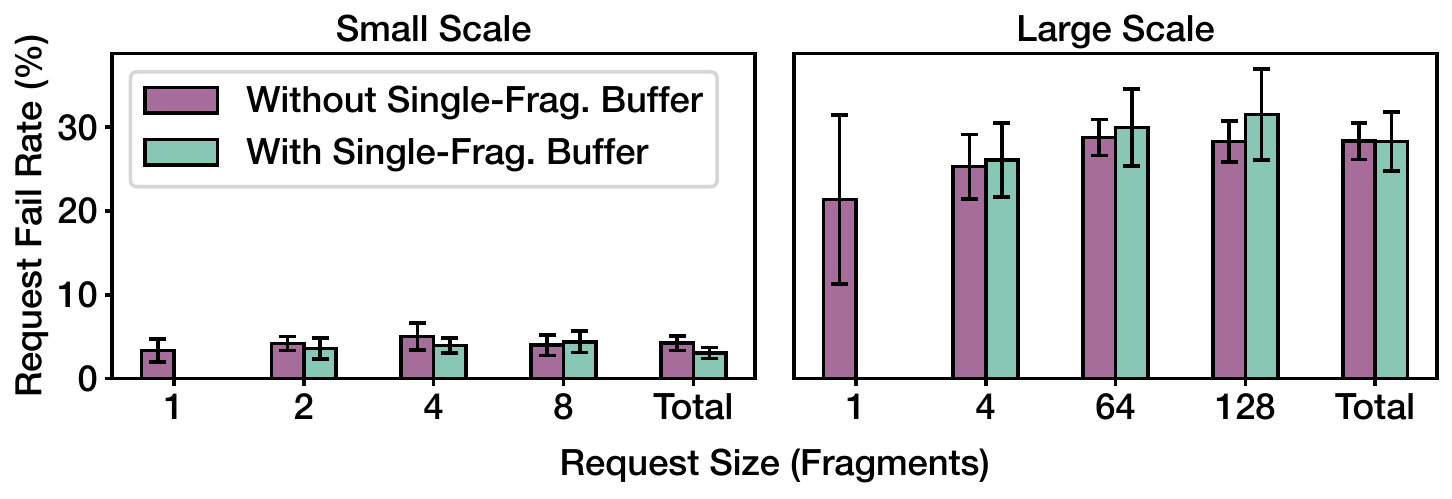}
  \caption{Request fail rates with and without a single fragment buffer.}
  \label{fig:reassembly-buffer-with-single}
\end{figure}

Next we ask \emph{should single-fragment requests be buffered along with larger requests?}
Serverless workloads exhibit high request size variability (cf. Table~\ref{tab:acc-functions}).
Small single-fragment requests do not require reassembly. However, if forced to share buffers with larger requests that can occupy reassembly buffers for longer periods, they can suffer higher response latency, and the potential for request drops.
Creating a distinct buffer for single-fragment requests allows them to be queued without interference by larger requests, ensuring accelerators can service them in the gaps between larger requests being reassembled.
Figure~\ref{fig:reassembly-buffer-with-single} shows that adding a single-fragment buffer completely eliminates failures for single-fragment requests, solving request-level HoL blocking while still maintaining competitive request fail rates for multi-fragment requests.

\begin{figure}[t]
  \centering
  \includegraphics[width=1.0\linewidth]{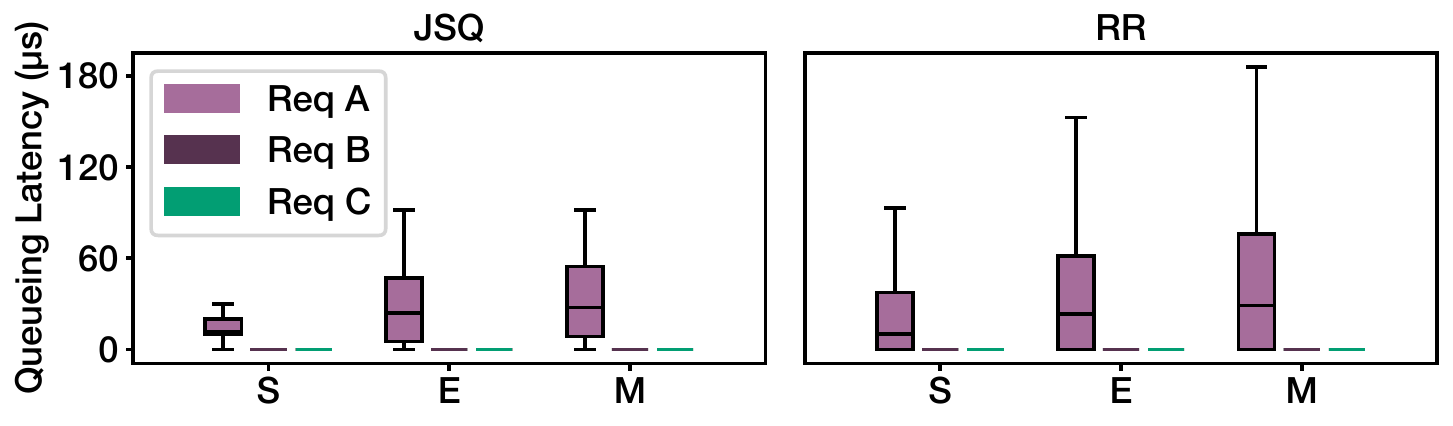}
  \caption{Queueing latency for accelerators A, B, and C under RR and JSQ for three request-size distributions: S (single-fragment dominant), E (even mix), M (multi-fragment dominant).}
  \label{fig:function-queue-policy}
\end{figure}
\subsection{Allocating Requests to Buffers}\label{sec:simreqbuffalloc}

We now ask \emph{how should a new incoming request be allocated to a specific reassembly buffer?} A naive round robin allocation between all buffers (Naive RR) would be problematic since buffers already allocated to an incomplete request are ineligible.
Hence, \offrac adopts a simple variant, where scheduling is performed only among eligible buffers (Eligible RR).
While more fine-grained strategies exist, like joining the shortest available buffer (JSB), \offrac favors Eligible RR for its hardware simplicity: it requires only a flag bit per buffer, whereas JSB would involve more complex comparisons between dynamic buffer states.
Moreover, our experiments (details omitted for brevity) demonstrate that Eligible RR achieves availability comparable to JSB.

\subsection{Allocating Requests to Multiple Accelerator Instances}
% \yzynote{I changed it from " Load Balancing for Accelerator Scaling" to this, which aligin with the above one}
% \kinote{I read it and it seems fine to me}
\label{sec:simloadbalanceaccel}

Even with high accelerator utilization, an accelerator with long execution time and a high request volume can be over-saturated.
To support scaling, \offrac supports hosting multiple instances of the same accelerator.
Ideally, optimal buffer utilization could be achieved by sharing a single queue among multiple accelerator instances of the same type, allowing completed requests to be load-balanced across instances.
However, this approach limits the framework's flexibility, as hardware queues must be pre-allocated, creating a tight coupling between accelerator types and queue instantiation.
To avoid this constraint, \offrac decouples queues from accelerator types and maintains distinct request queues for each accelerator slot.
Hence we ask \emph{how should assembled requests be allocated to multiple queues for the same accelerator type?}

We simulate two policies: Round Robin (RR) and Join Shortest Queue (JSQ), which selects the accelerator queue with the least queued data.
We simulate a workload with 95\% of requests for accelerator A and instantiate 3 instances of accelerator A and run three request size distributions (cf. Appendix~\ref{apx:simulator}, Table~\ref{tab:simulator_setup}).

Figure~\ref{fig:function-queue-policy} shows the queueing latency distribution for accelerators A, B, and C.
While simple RR achieves comparable median latency to JSQ, it has higher worst-case latency and higher tail latency.
This stems from RR assigning complete requests blindly in a fixed order, without considering request size, whereas JSQ accounts for the actual queue load.
In \offrac, we adopt RR for its hardware simplicity, accepting its tolerable trade-off in tail performance in exchange for efficient and practical implementation.

\subsection{On-Chip Buffering}
A key design decision in \offrac is to use on-chip memory to implement the reassembly buffers and accelerator queues.
This significantly reduces the overhead of reassembly and accelerator invocation, but also means the sizes of these buffers are constrained by limited on-chip memory.
Modern FPGAs offer 10s of MB of on-chip memory (about 9MB of BRAM and 35MB of URAM in the case of the AMD Alveo U280 used in our prototype~\cite{U280_BRAM}).
However, only a fraction of this (a few MBs) is available to \offrac, depending on how much of the FPGA is allocated to function acceleration.
At the highest level, if an accelerator has an execution time of $t_{\mathit{Acc}}$ and the input bandwidth is $\mathit{BW}$, then the buffers in aggregate should be able to buffer a data size of $\mathit{BW}\times t_{\mathit{Acc}}$.
Assuming $t_{\mathit{Acc}}$ of 100~$\mu$s and $\mathit{BW}$ of 100~Gbps, that equates to 1.25 MB of buffer space.

In \offrac, we can use more resources for the reassembly buffers or the accelerator queues.
Shorter accelerator queues mean complete requests may stay in reassembly buffers longer, causing fewer of them to be eligible, potentially resulting in request failures.
Longer accelerator queues can result in underutilized queue space when fewer accelerators are active or being requested.
Considering the ability to dynamically modify accelerators hosted in slots to balance incoming traffic demands, we choose to ensure accelerator queues can host multiple complete requests, while we size reassembly buffers to accommodate the size of the largest request plus some additional margin.
\subsection{Accelerator Virtualization}
To support accelerator virtualization, \offrac uses Partial Reconfiguration~\cite{amddfx} to replace the accelerator loaded in a slot at runtime.
This process is managed by a reconfiguration controller that coordinates slot reconfiguration on behalf of the control plane.
As illustrated in Figure~\ref{fig:architecture-design}, when a reconfiguration request arrives, the controller first waits until the accelerator currently assigned to the target slot becomes idle.
It then prevents the request selector from dispatching new requests to that slot, asserts the decoupling signals, and reconfigures the slot by streaming bitstream for the requested accelerator to the Internal Configuration Access Port (ICAP)~\cite{vipin2018fpga}.
Once reconfiguration succeeds, the controller updates the Selector's accelerator-to-slot allocation table so that subsequent requests for the newly loaded accelerator are dispatched to that slot.
It then de-asserts the decouplers and reports the reconfiguration status back to the control plane.
If an error occurs during reconfiguration, the slot remains decoupled and the control plane is notified of the failure.
This allows \offrac to support transparent accelerator hot-swapping while requests targeting other slots continue to execute without interruption.

\subsection{Putting It All Together}

\begin{figure}[t]
\centering
\includegraphics[width=0.95\columnwidth]{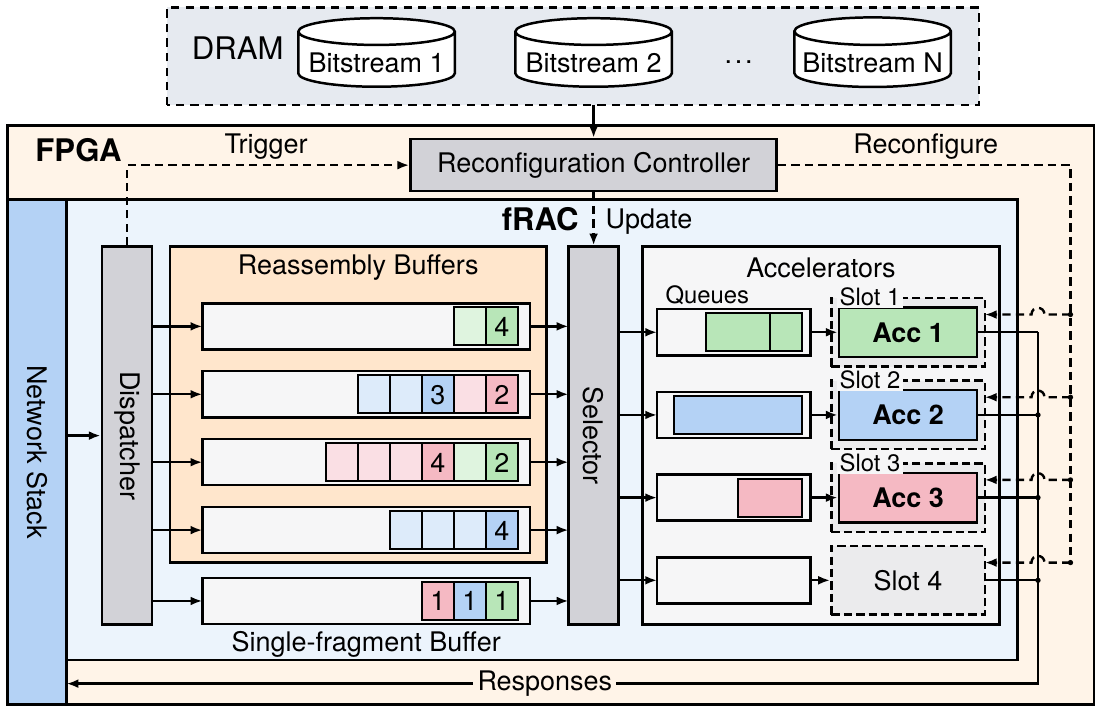}
\caption{\offrac design showing packet payloads passed to the Dispatcher as fragments which are then reassembled in buffers based on information in the request headers. Assembled requests are then passed to the relevant accelerator and responses returned to the network stack. Accelerators can be replaced by the Reconfiguration Controller.}
\label{fig:architecture-design}
\end{figure}

The design of \offrac builds atop an established network transport layer, leveraging ordered delivery of packet payloads to allow reassembly of application-level requests and accelerator invocation at that granularity.
Reassembly buffers combine request fragments into complete requests, which are accompanied by required invocation parameters.
These buffers are agnostic to the requesting client, the requested accelerator, and request size to provide high utilization in a dynamic workload scenario.
A separate buffer for small single-fragment requests ensures they are serviced reliably even in the presence of larger requests. The Dispatcher implements the buffer joining policies outlined above, which have minimal overhead.
Furthermore, the Reconfiguration Controller enables dynamic swapping of accelerators via Partial Reconfiguration, allowing the system to adapt to changing workloads at runtime without interrupting the network path.
The complete \offrac architecture is shown in Figure~\ref{fig:architecture-design}.

\section{Prototype Implementation}\label{sec:prototype}

Our open-source\footnote{\offrac codebase: \url{https://github.com/accl-kaust/frac}} hardware prototype of \offrac is built on an AMD Alveo U280 FPGA with 100G network interfaces.
The full request path is implemented within the FPGA. Packets enter through the 100G Ethernet interface, pass through the hardware TCP stack, are assembled by \offrac into complete requests, dispatched to the target accelerator, and returned as responses through the same network path.

\begingroup\emergencystretch=1.5em
\noindent \textbf{Network Stack.}
We implement our prototype in Verilog, building atop the open-source TCP/IP stack in~\cite{he2021easynet} for network transport.
Since \offrac is a hostless framework, all transport-layer functionality is implemented directly in hardware.
The hardware transport stack manages the complete TCP connection lifecycle, including establishment, per-connection transport state, reliable delivery, and teardown. No host participates in connection management.
We choose this TCP/IP stack because it is reliable and easy to extend with custom logic.
However, \offrac is transport-agnostic, meaning it can integrate with any protocol that guarantees packet ordering.
Our design, depicted in Figure~\ref{fig:architecture-design}, is implemented in the \textit{user kernel} portion of the network stack design.
TCP headers of individual packets are processed by the TCP/IP stack, which extracts the connection ID and passes it along with packet payloads as the fragments to \offrac.\par\endgroup

\begin{figure}[t]
  \centering
  \includegraphics[width=0.95\columnwidth]{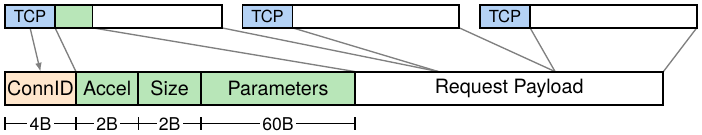}
  \caption{Request format showing the payloads of multiple TCP packets assembled into a request with its request header.}
  \label{fig:Request-format}
\end{figure}

\noindent \textbf{Request Format.}
Figure~\ref{fig:Request-format} illustrates the per-request header and its placement within TCP segments. We define a small 64B custom \emph{request header} that is embedded in the first fragment in a request.
This header identifies the \texttt{Accelerator} (2B) and the request \texttt{Size} (2B), plus 60B as the \texttt{Parameters} field carrying accelerator-specific parameters, e.g., the value $K$ for the Top-K accelerator. \texttt{Parameters} is opaque to \offrac, which passes it to the accelerator with the payload data.
The size of this field (60B) is standardized to simplify our implementation, but other designs are possible.

\noindent \textbf{Reassembly Buffers.}
The reassembly buffers use the connection ID and \texttt{Size} in the first fragment of a request to determine how many fragments to assemble into a complete request.
The reassembly buffers are implemented as First-In-First-Out buffers (FIFOs) utilizing on-chip BRAM. This allows for minimal buffering overhead and low latency.
The Dispatcher implements the buffer allocation policy to assign incoming fragments to reassembly buffers.
The Selector maintains a table of the currently instantiated accelerator in each slot and monitors all reassembly buffers in every cycle to identify completely reassembled requests that are then forwarded to the corresponding accelerator queues.
Requests are passed to the correct accelerator based on the \texttt{Accelerator} field in the header, along with the \texttt{Parameters}.

Our prototype implements four 4096$\times$512b (0.25MB) reassembly buffers and an additional 16384$\times$512b (1MB) single-fragment buffer.
Each accelerator queue is 1024$\times$512b (64KB), which significantly exceeds the largest request payload we support.
These buffers can be increased in size in a full deployment as modern FPGAs have more available on-chip memory in the 10s of MBs, especially when including URAM resources. While some of these must be retained for accelerator implementation, there is ample availability for buffers an order of magnitude larger than those in this proof of concept.

\noindent\textbf{Accelerators.} We implement a variety of accelerators (A1--A4 in Appendix~\ref{apx:resource-tables}, Table~\ref{tab:pr_func}), and consider additional accelerators from the literature (A5--A7 in Table~\ref{tab:pr_func}) when reasoning about partial reconfiguration and slot sizing.

\noindent \textbullet~\emph{Top-K (A1)}: processes a window of data, sorting and returning the highest $K$ integers, with $K$ being a dynamically set parameter per request (implemented based on~\cite{Istvan2014Histograms}). This operation is commonly used in data analytics for ranking tasks, such as extracting the most significant keywords or trending terms, as demonstrated in~\cite{Truica2017T2K2}.

\noindent \textbullet~\emph{Logit Transform (A2)}: applies a logit transformation over a floating point tensor block, performing subtraction, division, and a logarithm operation in a pipeline. This transformation plays a vital role in pre-processing within Deep Learning Recommendation Models (DLRMs), converting raw input features into normalized probability distributions for personalized recommendation predictions~\cite{Naumov2019DLRM}.

\noindent \textbullet~\emph{Min-Max Normalization (A3)}: normalizes a floating point tensor block, scaling each element by the range of the block with pipelined range determination, subtraction, and division. This is a crucial pre-processing step within DLRMs, standardizing raw input features into a consistent [0, 1] range to ensure uniform scaling across diverse data types \cite{Naumov2019DLRM}.

\noindent \textbullet~\emph{CNN (A4)}: A quantized neural network accelerator designed for image classification using multiple convolutional layers (filters from 8 to 32 elements), max pooling, and dense layers, trained on the SVHN dataset via hls4ml~\cite{fastml_hls4ml}. It processes 24KB input images (64x64 48-bit/pixel) and produces a one-hot encoded 10-class output. This represents a computationally intensive application requiring request reassembly in all cases and having a much longer execution time.

\noindent \textbf{Partial Reconfiguration.}
\label{sec:pr-analysis}
We must first define a set of Partially Reconfigurable Regions (PRRs) or slots, which must be spatially arranged subject to constraints relating to the underlying FPGA resources.
To maximize the area available to accelerators, we must constrain the network stack and \offrac infrastructure portions to parts of the FPGA that include the required I/O, while leaving as much area available to accelerator slots as possible.
This is highly challenging in a design that must achieve timing closure to function correctly with the I/O interfaces.

\begin{table}[t]
\small
\centering
  \caption{PR slots and reconfiguration times (ms).}
  \label{tab:pr_regions}
  \begin{tabular*}{\columnwidth}{@{\extracolsep{\fill}}crrrrrc@{}}
    \toprule
    \textbf{Slot} & \textbf{LUT} & \textbf{FF} & \textbf{BRAM} & \textbf{DSP} & \textbf{URAM} & \textbf{Time}  \\
    \midrule
    1 & 213k & 429k & 288 & 1440 & 192 & 6.90\\
    2 & 120k & 245k & 168 & 912 & 128 & 4.12\\
    3 & 120k & 245k & 168 & 912 & 128 & 4.12\\
    4 & 319k & 641k & 480 & 2304 & 256 & 9.92\\
    \bottomrule
    \multicolumn{7}{l}{\textit{Slots 1,4: all accelerators; Slots 2,3: A1--A3, A5--A7.}}
  \end{tabular*}
\end{table}

\begin{figure}[t]
  \centering
  \begin{minipage}[t]{0.48\columnwidth}
    \centering
    \rotatebox{90}{%
      \includegraphics[width=0.72\linewidth]{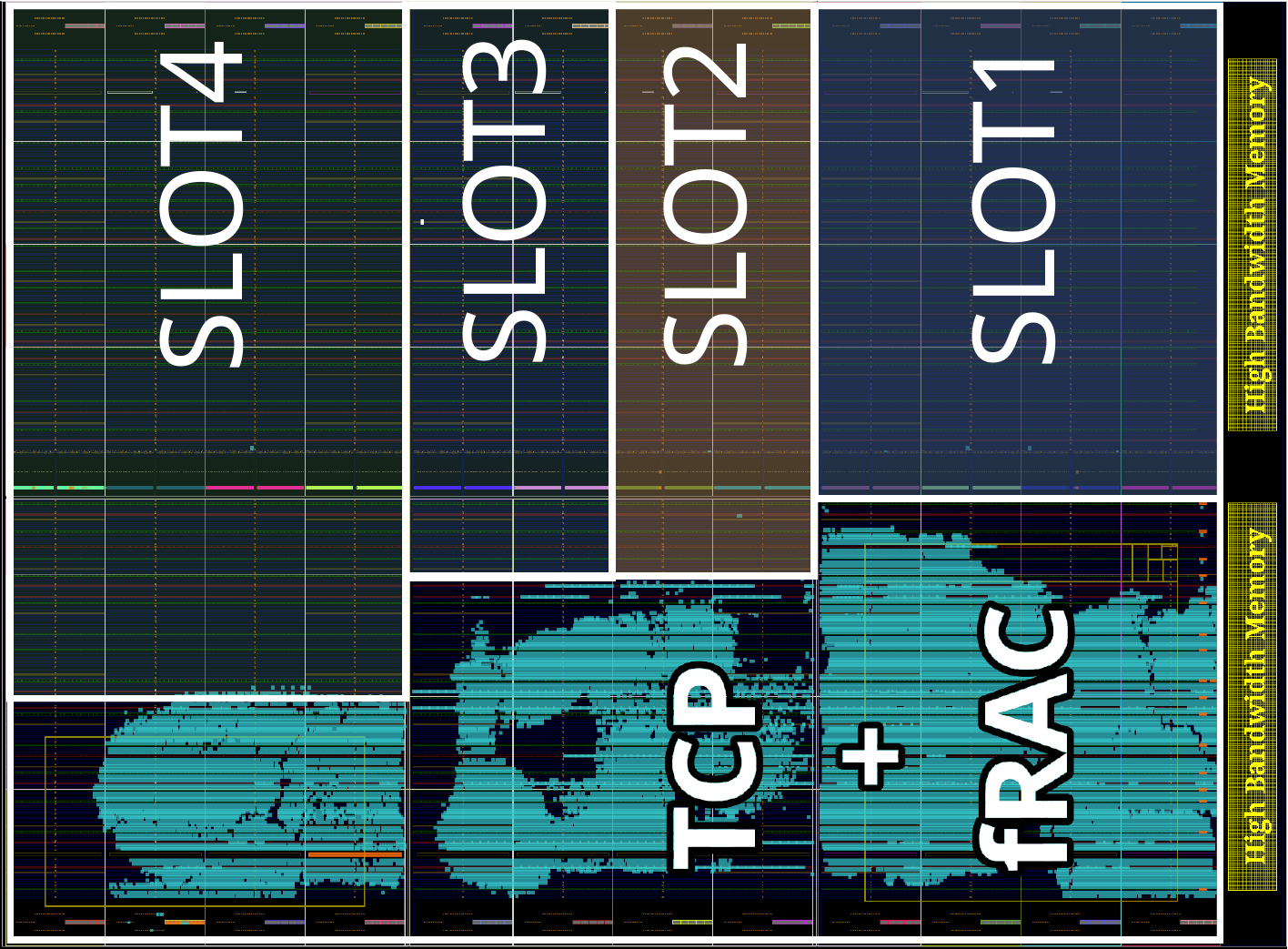}%
    }

    (a) 4-slot config.
  \end{minipage}\hfill
  \begin{minipage}[t]{0.48\columnwidth}
    \centering
    \rotatebox{90}{%
      \includegraphics[width=0.72\linewidth]{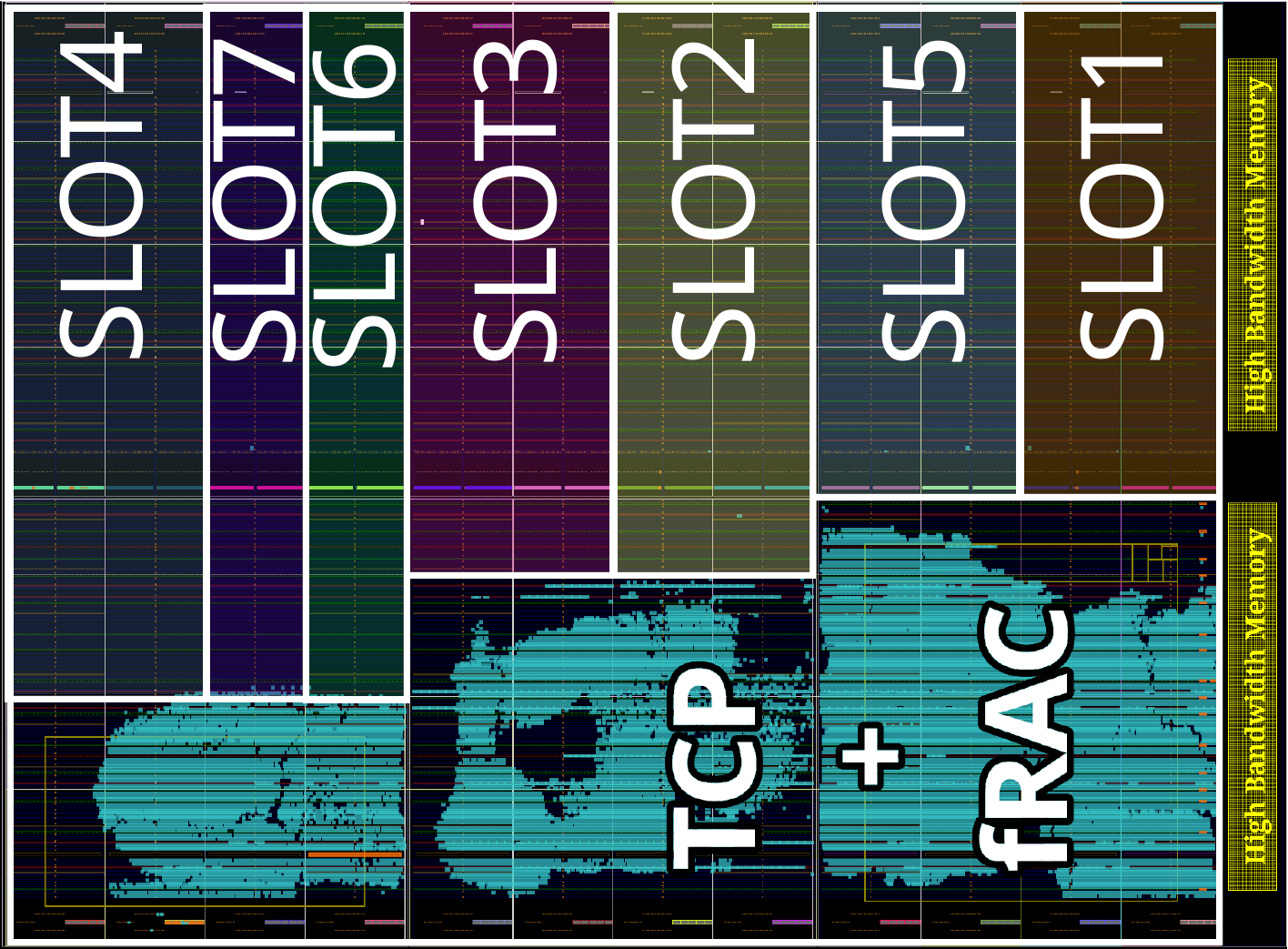}%
    }

    (b) 7-slot config.
  \end{minipage}
  \caption{Accelerator slot layouts on AMD Alveo U280.}
  \label{fig:offrac_floorplans}
\end{figure}

Since our design is hostless, we are able to remove hardware devoted to host PCIe communication, eliminating the XRT shell~\cite{amdxrt} and freeing additional FPGA area for accelerators. This optimization reduces the stack's resource usage by about 45\% compared to the original design.
Table~\ref{tab:resourceusage} presents the resource usage of the original and optimized TCP/IP stack as well as \offrac.

\begin{table}[t]
  \centering
  \footnotesize
  \caption{FPGA resources consumed by \offrac.}
  \label{tab:resourceusage}
  \setlength{\tabcolsep}{3pt}
  \begin{tabular*}{\columnwidth}{@{\extracolsep{\fill}}lccccc@{}}
    \toprule
    \textbf{Component} & \textbf{LUT} & \textbf{FF} & \textbf{BRAM} & \textbf{URAM} & \textbf{DSP}\\
    \midrule
    \shortstack{Original\\TCP/IP Stack} & \shortstack{260.4k \\(19.99\%)} & \shortstack{454.1k\\(17.43\%)} & \shortstack{443\\ (21.97\%)} & \shortstack{0 \\(0\%)} & \shortstack{0 \\(0\%)} \\ \midrule
    \shortstack{Optimized\\TCP/IP Stack} & \shortstack{150.5k \\(11.5\%)} & \shortstack{130.2k \\(5.0\%)} & \shortstack{200.5 \\(10.0\%)} & \shortstack{16 \\(1.7\%)} & \shortstack{0 \\(0\%)} \\\midrule
    \offrac Layer& \shortstack{5.1k \\(0.4\%)} & \shortstack{25.4k \\(1.0\%)} & \shortstack{250 \\(12.4\%)} & \shortstack{40 \\(4.2\%)} & \shortstack{0 \\(0\%)} \\
    \bottomrule
  \end{tabular*}
\end{table}

The buffer allocations (including reassembly buffers and single-fragment buffers) are aligned with the size mentioned above.
Figure~\ref{fig:offrac_floorplans} shows two example floorplans we designed with 4 or 7 accelerator slots to show slot configurations for different accelerators with different reconfiguration times, while respecting spatial layout constraints.
Table~\ref{tab:pr_regions} shows the resources available in each of the slots for the 4-slot configuration.
Cross-referencing against the accelerators in Table~\ref{tab:pr_func} in Appendix~\ref{apx:resource-tables}, we see that this arrangement can support a wide range of different accelerators being instantiated at the same time, including more complex accelerators than we built for this prototype.

To enable fast accelerator switching, \offrac uses the high-throughput Internal Configuration Access Port (ICAP)~\cite{vipin2018fpga}.
When an accelerator is loaded, the Selector's slot allocation table is updated.
We measured actual reconfiguration times, reported in Table~\ref{tab:pr_regions}, by issuing a request to reconfiguration controller and recording the time required for the controller to stream the corresponding accelerator bitstream from off-chip HBM to ICAP.
The measured times are roughly 1--2~ms higher than theoretical estimates based on ICAP bandwidth, due to occasional stalls introduced while reading the bitstream.
Compared to the 100ms to 1s cold-start delays commonly seen in software-based systems~\cite{noauthor_operating_2021}, \offrac can achieve reconfiguration in just a few milliseconds.
This design could be extended further using nested partial reconfiguration~\cite{amdnesteddfx} to allow the number and sizes of slots to be modified at runtime.

% \kinote{remove this para}
%\offrac uses partial reconfiguration to load accelerators on demand. As illustrated in Figure~\ref{fig:architecture-design}, when a client issues a request for an accelerator that is not currently loaded on the FPGA, the autoscaler triggers a reconfiguration command to the node agent, which selects an available slot and reconfigures it with the required accelerator. The Selector's slot--accelerator mapping table is updated to reflect the new configuration, and subsequent requests are routed to the newly instantiated accelerator. This enables transparent accelerator hot-swapping while allowing ongoing requests targeting other accelerators to proceed without interruption.

\noindent\textbf{Integrating New Accelerators.}
\label{apx:newaccel}
A key strength of \offrac is that integrating new accelerators is straightforward.
A 512-bit AXI-Stream interface is provided for data input and output.
A simple wrapper module can translate these streams into the required AXI-Stream data width of an arbitrary accelerator through FIFOs.
Another 32-bit AXI-Stream interface is provided for metadata input and output.

The accelerator is passed the 64B request header as the first piece of data, followed by the remainder of the payload.
Depending on the specific design of the accelerator, it may not require any of this information, in which case it can process the following stream.
It may use the \textsf{Size} field to configure internal execution, such as the number of iterations of a computation to perform,
and it may use the custom-defined \textsf{Parameters} field to configure other datapath options, such as the value of $K$ for a Top-K accelerator,
the datatype for an arithmetic kernel, the matrix dimensions for a matrix multiplication, or the kernel for an image filter.
The specification of these must, of course, be communicated to the clients so they can issue valid requests.
The accelerator produces its output data stream and must indicate the size of the response on the last beat using the metadata stream.
Some streaming accelerators may require additional logic to flush the pipeline after the completion of input ingestion to fit the run-to-completion model for a single request.
The wrapper ensures the connection ID is passed with the response to route to the correct client.

Consider a CNN accelerator generated with hls4ml~\cite{fastml_hls4ml}.
Both input and output interfaces use AXI-Stream ready/\allowbreak valid handshakes.
It ingests one 48-bit pixel every 223 cycles due to resource-shared pipelines. After reading 4096 inputs (64$\times$64 pixels), it produces the classification output 913{,}408 clock cycles after the first input.
To integrate it into \offrac, we first assign it an accelerator ID to be used by the Selector.
We wrap the module with a FIFO-based adapter that bridges the 512-bit \offrac stream to the 48-bit interface of the IP.
A counter in the wrapper ensures sufficient input padding is applied to flush the outputs for each complete input.
Simpler accelerators such as Logit Transform and Min-Max Normalization exhibit more straightforward behavior and do not require flushing.
As long as an accelerator exposes a streaming interface and its execution semantics are understood, it can easily be wrapped to integrate into the \offrac framework without changes to its core logic.
Notably, this CNN accelerator embeds all weights and biases in the logic using on-chip memory. While the accelerators we implemented do not use off-chip memory, connectivity to external HBM channels through AXI interconnect is supported.
Table~\ref{tab:pr_func} in Appendix~\ref{apx:resource-tables} shows the hardware implementation results of the accelerators we implemented (A1--A4) as well as additional accelerators from the literature (A5--A7), which we use only as comparative reference designs.

\offrac also provides abstract shells for the layouts shown in Figure~\ref{fig:offrac_floorplans}, allowing new accelerators to be synthesized and implemented without rebuilding the full \offrac stack.
This can save hours of vendor-tool compilation time during accelerator development.
In contrast, under a typical PR flow, adding a new accelerator would require regenerating all accelerator bitstreams and updating the bitstream database stored in local memory.
The Abstract Shell flow~\cite{amdabstractshell} avoids this restriction, allowing a new accelerator to be implemented without invalidating the existing static design or previously generated accelerator bitstreams.
This separation also gives cloud infrastructure maintainers a practical way to support accelerator development, since they can provide the abstract shells without exposing the complete \offrac design.

\section{Evaluation}
\label{sec:evaluation}

% \addnote{We do not compare against mainstream serverless platforms (e.g., OpenFaaS~\cite{openfaas} and AWS Lambda~\cite{aws_lambda})
% their end-to-end invocation overhead is typically in the millisecond range\cite{Jia2021Nightcore} dominated by software control-plane and platform costs,
% Such systems would therefore provide an unchallenging baseline for \offrac's microsecond-scale.
% Similar problem exists for host-controlled FPGA solutions with F3 \cite{mainas_f3_2025} an overhead of 10.8 $\mu$s solely for FPGA API call,
% over an order of magnitude higher than \offrac's 0.16 $\mu$s shown in Figure~\ref{fig:throughput_latency}.
% Also lambda NIC \cite{choi2020lambda} shows a median latency exceeding 50 $\mu$s for the simplest web workload, far higher than \offrac's latencies for substantially more complex logic workloads.
% We adopt a more demanding baseline by comparing against a highly optimized DPDK-based Libtpa server (and a DPU implementation)
% that removes most OS and framework overheads yet still runs on general-purpose CPUs.
% }

We explore whether \offrac:
\begin{enumerate*}[label=(\roman*)]
    \item \textit{exhibits minimal overhead over the transport layer it builds upon, and sustains low latency under high throughput conditions;}
    \item \textit{handles mixed accelerator requests while ensuring performance isolation;}
    \item \textit{provides efficient reassembly to increase accelerator throughput;}
    \item \textit{outperforms optimized software implementations in scalability and delivers more consistent latency;}
    \item \textit{maintains stable performance under a real-world serverless trace.}
\end{enumerate*}
A full evaluation artifact for these experiments is available.\footnote{\url{https://zenodo.org/records/23200153}}

\begin{figure*}[tb]
  \centering
  \begin{minipage}[t]{0.32\textwidth}
    \centering
     \includegraphics[width=\linewidth]{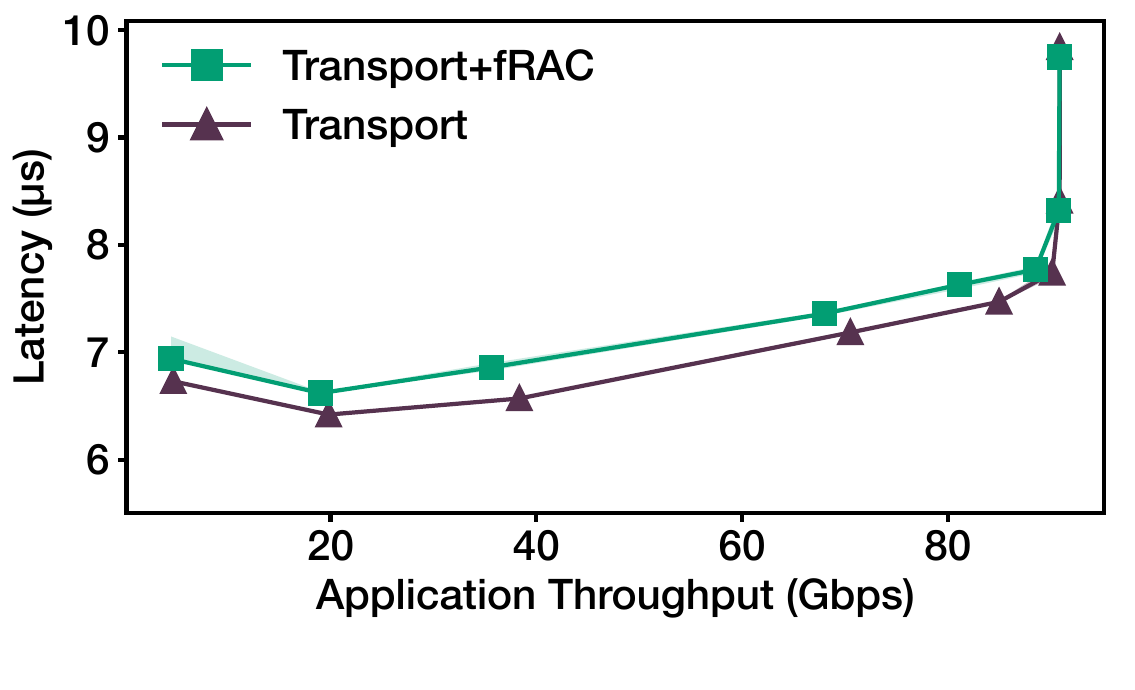}
    \caption{Median latency vs application throughput with/without \offrac (shaded 25--75th pct.).}
    \label{fig:throughput_latency}
  \end{minipage}
  \hfill
  \begin{minipage}[t]{0.32\textwidth}
    \centering
    \includegraphics[width=\linewidth]{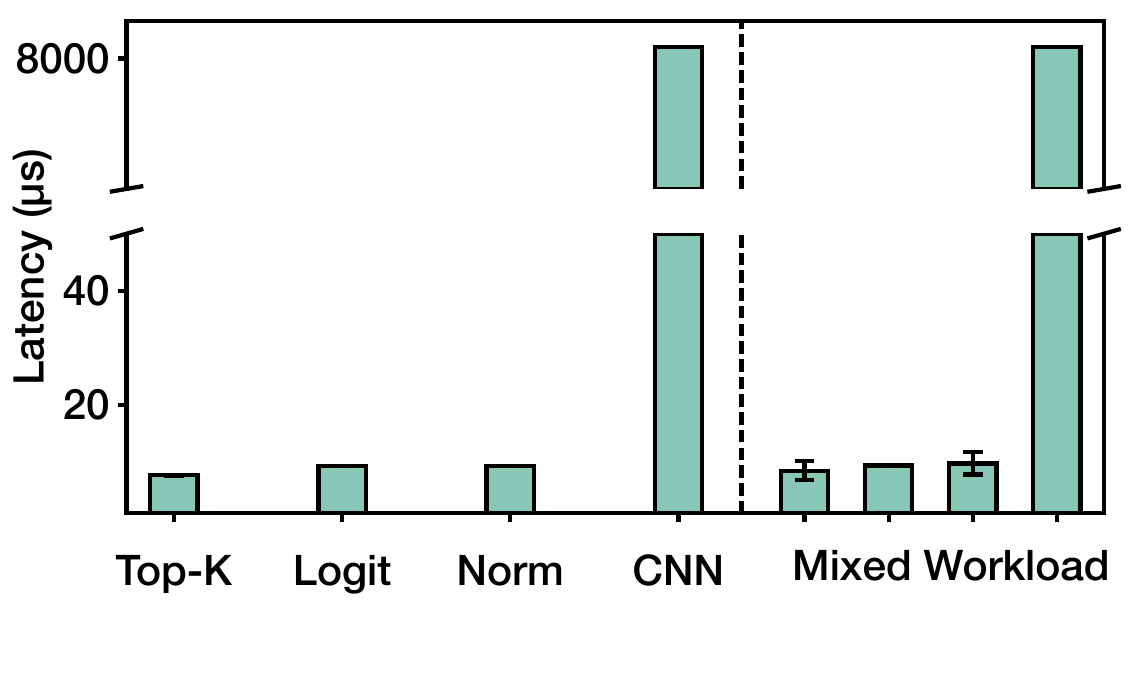}
    \caption{Comparison of latency between isolated baselines and a workload with mixed requests.}
    \label{fig:performance_isolation}
  \end{minipage}
  \hfill
  \begin{minipage}[t]{0.32\textwidth}
    \centering
    \includegraphics[width=\linewidth]{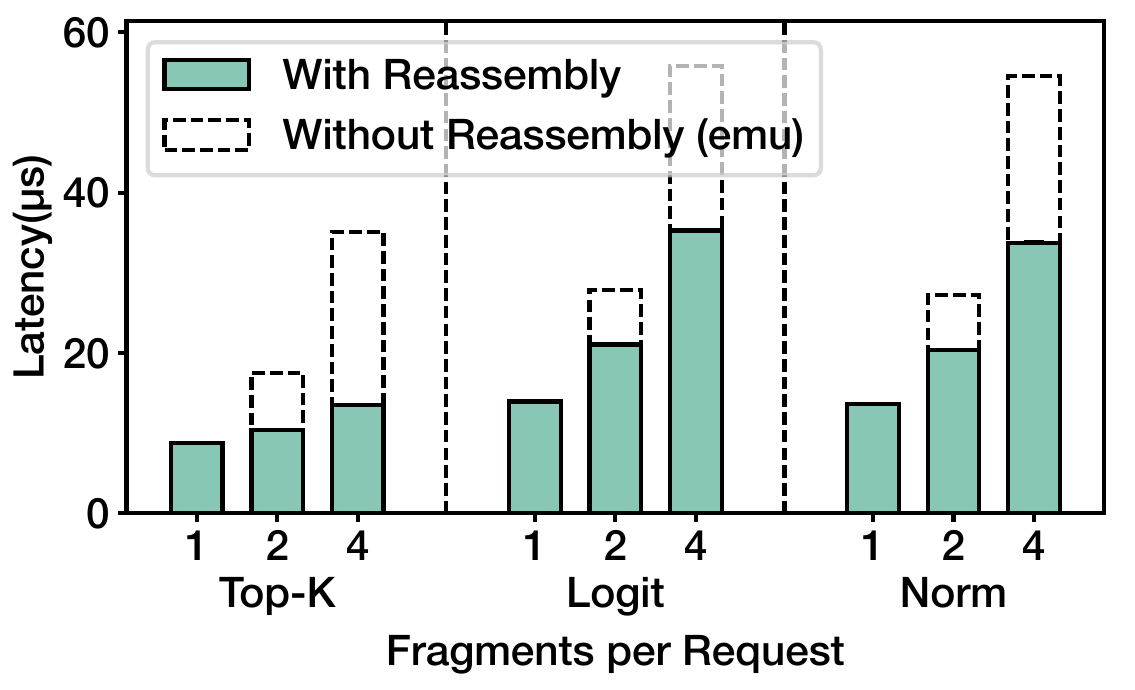}
    \caption{Impact of request size on latency.}
    \label{fig:performance_reassemble}
  \end{minipage}
\end{figure*}

\noindent \textbf{Choice of Baselines.}
We do not include mainstream serverless platforms such as OpenFaaS~\cite{openfaas} and AWS Lambda~\cite{aws_lambda} in our quantitative comparison as their end-to-end invocation latencies are typically in the millisecond range~\cite{Jia2021Nightcore}, dominated by software platform overheads.
Similarly, host-controlled FPGA frameworks have high overhead. Molecule~\cite{Dong2022Serverless} incurs 50--100~$\mu$s to DMA just 4KB between CPU and FPGA. F3's FPGA API call overhead is 10.8~$\mu$s~\cite{mainas_f3_2025}, over an order of magnitude higher than \offrac's 0.16~$\mu$s framework overhead in Figure~\ref{fig:throughput_latency}.
$\lambda$-NIC reports median latencies above 50~$\mu$s even for simple workloads~\cite{choi2020lambda}.
We adopt more demanding baselines by comparing against a highly optimized DPDK-based Libtpa server and a DPU implementation.

\noindent \textbf{Testbed and Experimental Setup.}
Clients run on a machine with AMD EPYC 7763 64-core processor, 512 GB DDR4 RAM, NVIDIA ConnectX-6 NIC, running Ubuntu 20.04.
The client uses the Libtpa TCP stack based on DPDK (see Appendix~\ref{apx:benchmarking}). We heavily adapted Libtpa~\cite{libtpa} for our evaluations.
Each client is pinned to a hardware thread, all within the same NUMA node connected to the NIC.
\offrac is deployed on an AMD Alveo U280 FPGA.
The FPGA receives requests from its QSFP network interfaces, processes them, and returns responses back to the client directly.
For the CPU comparison, we use another machine with identical configuration running a DPDK-based server using Libtpa.
For the DPU baseline, we deploy an NVIDIA BlueField-3 SmartNIC~\cite{BlueField3} in off-path mode with identical configuration,
as off-path is the only mode that supports general-purpose application offloading and all function processing runs solely on the DPU Arm cores, with no involvement of the host CPU.
All interfaces are connected to an EdgeCore DCS810 switch at 100 Gbps, with an MTU configuration of 9000B.
All \offrac experiments in this section run on the Alveo U280 FPGA. Persistent TCP connections are established before measurement, so latencies exclude connection establishment and teardown. We use saturation experiments to expose throughput ceilings and queueing behavior, and trace replay to evaluate time-varying production workloads.

\noindent \textbf{Low Invocation Latency.} We first evaluate the latency overhead of the \offrac abstraction under increasing throughput.
We implement an echo function directly atop the underlying TCP stack, and as a function with the \offrac abstraction.
For this experiment only, the client and the FPGA are connected directly with a cable without a switch, so that the measured latency reflects the limit of the \offrac prototype itself rather than switch-induced delay.
We compare the latencies under increasing load with a varying number of clients.
Each client opens a TCP connection and transmits single-fragment requests with a 4096B payload for 30 seconds in a closed loop, meaning each client issues its next request only after receiving the response to its previous request, keeping at most one request outstanding.
The first 10 seconds of results are discarded. With 24 concurrent clients, we achieve approximately 90 Gbps of application throughput (i.e.,~considering only the payload).
The 24-client count is simply an artifact of the number of clients needed to reach maximum throughput in this high-load configuration.
%todo how many clients
As shown in Figure~\ref{fig:throughput_latency}, \offrac introduces minimal overhead, adding only about 0.16 $\mu$s compared to echoing data in the transport layer.
Even at peak load, overall latency remains below 8 $\mu$s, demonstrating excellent \emph{scalability} under saturating network conditions.

\noindent \textbf{Performance Isolation.} To demonstrate that \offrac suitably isolates requests for different accelerators, even when their execution times vary significantly (cf. Table~\ref{tab:acc-functions}  and Table~\ref{tab:pr_func}), we deploy four accelerators (Top-K, Logit Transform, Min-Max Normalization, and CNN) in separate slots.
We first conduct a single-client test for each accelerator to observe its baseline latency in isolation.
To evaluate performance isolation under contention, we next run a mixed workload with four concurrent clients each sending requests exclusively to one of the four for five seconds.
Logit Transform and Min-Max Normalization clients send single-fragment requests of 1024B, while CNN workload requests are six 4096B fragments, forming one input image.
Figure~\ref{fig:performance_isolation} shows that latency remains consistent for isolated and mixed workloads, confirming that performance isolation is maintained even when accelerators have different execution times.
Error bars may not be visible due to low variance.

\noindent \textbf{Accelerator Utilization.} We extend the previous experiment with clients sending requests of 1, 2, and 4 fragments (each 1024B)
in a closed loop to demonstrate the benefit of request reassembly.
We measure latency from last fragment sent to response received.
Run to completion means processing multi-fragment requests requires multiple invocations without reassembly.
Figure~\ref{fig:performance_reassemble} shows that, with reassembly, latency increases sub-linearly with request.
For Logit Transform and Min-Max Normalization, an assembled 2-fragment request takes 76\% of the time of processing its fragments separately, and a 4-fragment request 64\%.
For a lightweight workload like Top-K, where communication overhead exceeds processing time, the advantage is larger. Error bars may not be visible due to low variance.
\begin{figure*}[tb]
  \centering
  \includegraphics[width=\textwidth]{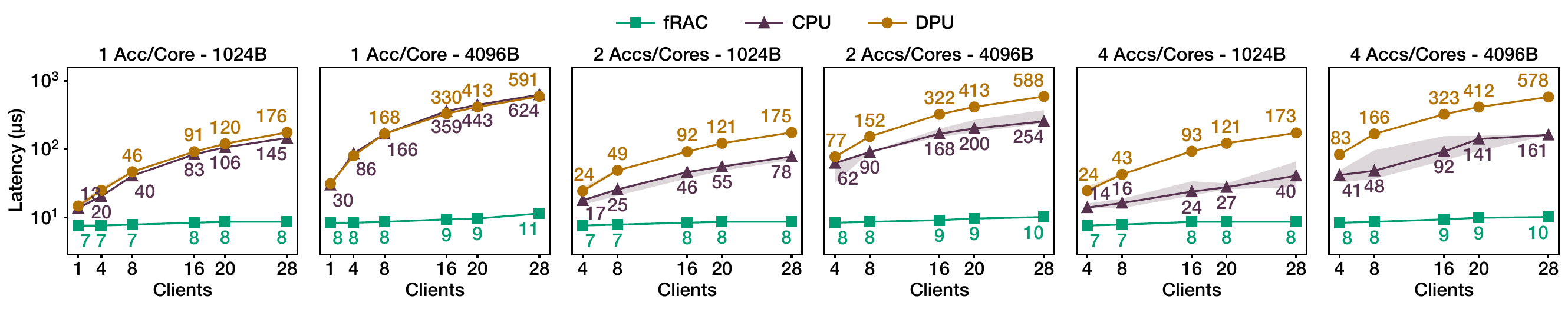}
  \caption{Latency comparison for the Top-K function on \offrac, CPU, and DPU, showing median, 25th, and 75th percentiles.}
  \label{fig:CPU_comparison}
\end{figure*}

\begin{figure*}[tb]
\centering
\begin{minipage}[b]{\columnwidth}
  \centering
  \includegraphics[width=\linewidth]{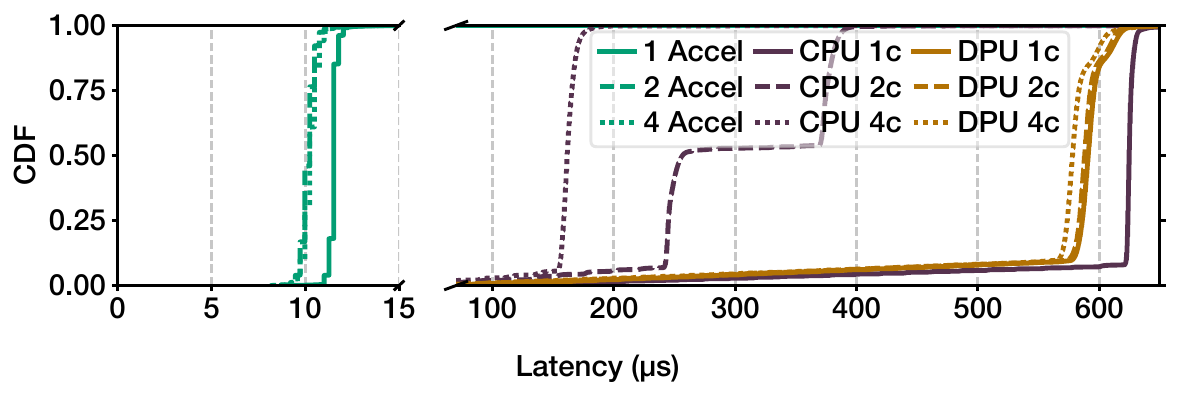}
  \caption{Latency CDF for saturating 4KB requests.}
  \label{fig:tail_cdf}
\end{minipage}
\hfill
\begin{minipage}[b]{\columnwidth}
  \centering
  \includegraphics[width=\linewidth]{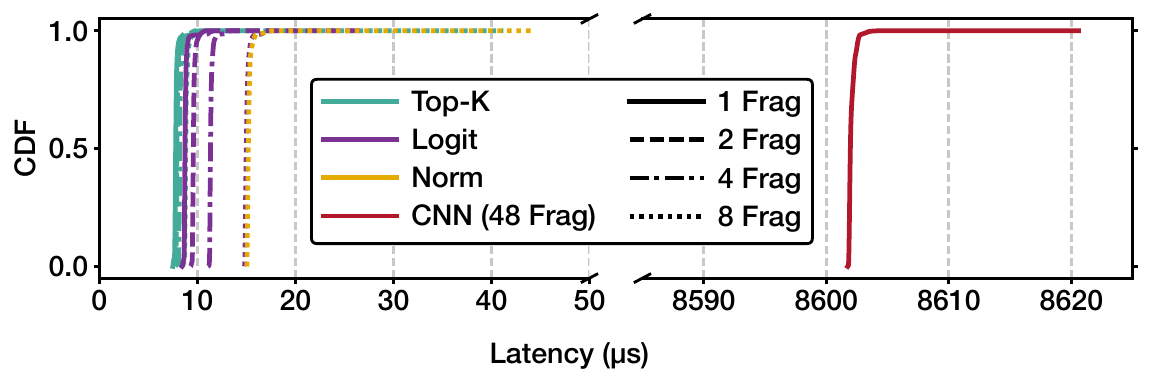}
  \caption{Latency CDF for the Azure trace replay.}
  \label{fig:azure_trace_cdf}
\end{minipage}
\end{figure*}

\noindent \textbf{Scalability Comparison with CPU and DPU.} To demonstrate that \offrac provides better scalability and more stable latency than function execution on CPU or DPU, we use high-performance DPDK Libtpa (Appendix~\ref{apx:benchmarking}) as a baseline.
DPDK is not a standard stack for serverless applications; most frameworks typically exhibit end-to-end latencies ranging from hundreds of $\mu$s to tens of ms~\cite{Jia2021Nightcore}, as demonstrated in host-managed offload~\cite{bacis2020blastfunction,mainas_f3_2025,Dong2022Serverless}.
However, this more demanding baseline highlights the performance gains achieved by eliminating host-side control in \offrac.
We instantiate \offrac with 1, 2, and 4 Top-K accelerators and compare against function invocation on server CPU and the Arm cores of the NVIDIA DPU, both using Libtpa (Appendix~\ref{apx:benchmarking}), pinned to 1, 2, and 4 CPU cores on each platform for fair comparison, ensuring equivalent parallel processing.
We scale the number of clients up to 28 to saturate network bandwidth (this is the minimum required to reach maximum DPDK throughput in our closed-loop setup, \offrac can scale to hundreds of clients).
Each client sends single-fragment requests of 1024B or 4096B.
While this workload does not exercise request reassembly, it serves to highlight latency differences among the platforms.
Figure~\ref{fig:CPU_comparison} shows that \offrac reduces latency by 46--95\% for 1024B requests and 73--98\% for 4096B requests compared to the CPU,
and 53--95\% and 73--98\% compared to the DPU.
Latency on the CPU and DPU increases linearly with the number of clients.
In contrast, \offrac latency remains nearly flat regardless of the number of clients and accelerators: for example, with a single accelerator and 4096B requests, latency rises only from 8 to 11~$\mu$s from 1 to 28 clients, while CPU latency grows from 30 to 624~$\mu$s.
Figure~\ref{fig:tail_cdf} shows \offrac's much lower tail latency under full load (28 clients).
We also ran the CNN workload on \offrac and the CPU server: each 24~KB request (six 4096B fragments) takes 8.6~ms on \offrac versus 40~ms with C-compiled TensorFlow on the CPU.

\noindent\textbf{Power Consumption.} Serving the same CNN client load, the CPU server and NIC consume 188--250~W from idle to full load (104--110~W from the CPU alone), while the hostless \offrac prototype consumes a total of 28--31~W, a 6--8$\times$ reduction in platform power.

% \removenote{We further compare \offrac to an NVIDIA BlueField-3 DPU~\cite{BlueField3} running the same Top-K function, pinned to 1, 2, and 4 cores to match the number of \offrac accelerators.
% Figure~\ref{fig:CPU_comparison} shows that 1, 2, and 4 DPU cores offer the same latency, comparable to a single CPU core.
% Figure~\ref{fig:tail_cdf} further shows that the DPU has much higher tail latency than \offrac.
% Across all configurations, both the CPU and DPU baselines consistently show substantially higher latency than \offrac.}

\noindent \textbf{Realistic Workload.}
We use the Azure Function Invocation Trace~\cite{zhang2021faster}, which reports two weeks of serverless function invocations.
We select the one-hour interval with the highest invocation density and replay it as a sequence of accelerator requests.
Pre-processing applied to the trace for our experiments is described in Appendix~\ref{apx:azure-trace}.
Figure~\ref{fig:azure_trace_cdf} shows the resulting latency CDF per accelerator (color) and request size in fragments (line style), demonstrating that \offrac maintains stable latency under a real-world workload.

\section{Comparison with Related Work}
\label{sec:related_work}
Table~\ref{tab:comparison} summarizes how \offrac compares to prior works across hardware transport support, request granularity, multi-tenancy, accelerator offload, and hostless deployment.
\input{comp-table.tex}
\noindent\textbf{Packet-level SmartNICs.} RingLeader~\cite{lin2023ringleader} and PANIC~\cite{lin2020panic} extend the Corundum~\cite{Corundum} FPGA NIC for intra-server orchestration and limited network function offload, respectively.
SuperNIC~\cite{lin2024supernic} offloads graphs of network tasks.
ClickNP~\cite{Li2016ClickNP} provides modular network function offload without protocol support.
All target datacenter traffic routing to host applications rather than application-level function acceleration.

\noindent\textbf{Application-Specific Processing.} High-Frequency Trading (HFT) accelerators on FPGA hardwire exchange-specific parsing and trading logic into the network datapath~\cite{leber2011hft}. They optimize a fixed application pipeline, whereas \offrac routes flexible requests between clients and accelerators.

\noindent\textbf{Transport Offload and Stream Processing.} TCP Offload Engines (TOEs) move connection management and in-order byte-stream reconstruction from the host CPU to the NIC~\cite{mogul2003tcp}. Pigasus~\cite{zhao2020pigasus} couples per-flow TCP reassembly to a fixed intrusion detection pipeline. In contrast, \offrac inserts a request-level hardware shim between the transport and accelerators. A shared, connection- and accelerator-agnostic buffer pool decouples data ingestion from execution, routing complete requests to virtualized, reconfigurable accelerators.

\noindent\textbf{Host-Managed FPGA Virtualization.} Frameworks managing data movement and reconfiguration in software~\cite{chen2014enabling,fahmy2015virtualized} incur significant overhead for networked applications, as seen in BlastFunction~\cite{bacis2020blastfunction}, Molecule~\cite{Dong2022Serverless}, F3~\cite{mainas_f3_2025}, and Harmonia~\cite{Li2025Harmonia}.
Coyote~\cite{korolija2020OS,Ramhorst2025Coyote} is a heavyweight OS-FPGA hybrid system providing transport support and reconfigurable accelerators.
Strega~\cite{maschi2024strega} extends EasyNet~\cite{he2021easynet} for request-level HTTP calls but remains host-managed and static.
RpcNIC~\cite{zhang_rpcnic_nodate} runs reconfigurable SmartNIC kernels but relies on the host CPU for control and scheduling.
Split gRPC~\cite{ramos2024splitgrpc} partitions RPC processing between the host and SmartNIC for isolation, but retains host-side execution and orchestration.

\noindent\textbf{Streaming and Request-level Processing.} Ens{\=o}~\cite{sadok2023enso} and nanoPU~\cite{Stephen2021NanoPU} reassemble packets for efficient NIC-to-software streaming without accelerating the workload itself.
Beehive~\cite{lim2024beehive} offloads transport protocols but lacks multi-tenancy and application-level abstractions.
NEBULA~\cite{sutherland_nebula_2020} employs a stateless request/reply protocol eliminating network/PCIe overheads, but lacks accelerator offload.
FlexDriver~\cite{FlexDriver2022Eran} supports request-level granularity but separates the network interface from accelerators, forcing data over PCIe.
Rain~\cite{ma2026rain} schedules microsecond-scale requests in-network via programmable switches, focusing on cluster-level load balancing rather than endpoint accelerator offload.
Arca~\cite{srivatsan2026arca} executes fine-grained serverless functions on CPUs using continuations rather than invoking hardware accelerators.

In contrast, \offrac invokes virtualized, runtime-recon\-fig\-urable accelerators on complete requests, entirely within the FPGA and without a host.

\section{Discussion and Conclusions}
\label{sec:conclusion}

We have proposed the Remote Accelerator Call (RAC) paradigm, which decouples network data transfer from accelerator control through a request-level invocation abstraction. Our \offrac prototype demonstrates that network-facing FPGAs can directly form, dispatch, and launch accelerator work, completely bypassing host-software bottlenecks. We validated that this hostless approach achieves microsecond-scale latency and high throughput, even under heavy load, outperforming host-managed approaches at significantly reduced power consumption.

\offrac's design is intentionally modular: the number and size of reassembly buffers and accelerator slots can be tuned to different deployments. The simple per-buffer signaling scheme allows the Dispatcher to change how buffers are shared without modifying accelerator logic. On the client side, a lightweight RPC-style library hides transport details, allowing applications to seamlessly leverage \offrac's hardware interface. We also demonstrate that \offrac can reconfigure accelerator slots within a few milliseconds, allowing the set of deployed accelerators to change at runtime to adapt to evolving demand.
We have demonstrated an extension to the request-level interface design to support chained function pipelines within and across \offrac nodes in~\cite{yang2026pipelines}, demonstrating scalability to more complex application graphs.

Exploring integration of \offrac with alternative transport protocols is a natural next step, enabling better coupled use of hardware buffer resources, and even lower latency.
Developing a control plane that uses live hardware queue signals to dynamically scale accelerator instances on demand while balancing of load across instances would further improve workload adaptation.
We also plan to extend \offrac to support more complex accelerators that rely on off-chip memory.
In the longer term, we envisage an \offrac-like invocation abstraction as a hardened interface in future network-attached devices.

\begin{acks}
This publication is based upon work supported by \grantsponsor{kaust-orfs}{King Abdullah University of Science and Technology (KAUST)}{} under Award No.~\grantnum{kaust-orfs}{ORFS-CRG11-2022-5017}. We thank Gianni Antichi and R.~Srikant for helpful discussions, and the anonymous reviewers for their comments.
\end{acks}

\clearpage
\appendix
\setlength{\intextsep}{4pt}
\begin{table}[b]
  % Accelerator colors match Acc 1/2/3 in Figure~\ref{fig:architecture-design}.
  \definecolor{accA}{HTML}{B8E6B8}
  \definecolor{accB}{HTML}{B4D2F5}
  \definecolor{accC}{HTML}{F5B9C3}
  \DeclareRobustCommand{\chip}[2]{{\setlength{\fboxsep}{0.4pt}\colorbox{#1}{\vphantom{0/}#2}}}
  \newcommand{\ca}{\chip{accA}}
  \newcommand{\cb}{\chip{accB}}
  \newcommand{\cc}{\chip{accC}}
  \caption{Simulation parameters for each experiment. Mix, exec.\ time, and instances are per accelerator (\ca{A}:\cb{B}:\cc{C}). PA: per-accelerator buffers; NRR/ERR: Naive/Eligible RR.}
  \label{tab:simulator_setup}
  \centering
  \footnotesize
  \newcommand{\cstack}[1]{$\vcenter{\hbox{\shortstack{#1}}}$}
  \newcommand{\chipstack}[1]{{\renewcommand{\arraystretch}{0.92}\begin{tabular}[c]{@{}c@{}}#1\end{tabular}}}
  \setlength{\tabcolsep}{2.5pt}
  \begin{tabular}{@{}lcccccccc@{}}
  \toprule
  & \multicolumn{3}{c}{\textbf{Generator}} & \multicolumn{2}{c}{\textbf{Buffers}} & \multicolumn{3}{c}{\textbf{Accelerators}}\\
  \cmidrule(lr){2-4}\cmidrule(lr){5-6}\cmidrule(lr){7-9}
  & \textbf{Cl.} & \textbf{Mix (\%)} & \textbf{Sizes} & \textbf{No.} & \textbf{Policy} & \textbf{Policy} & \textbf{Exec.\ ($\mu$s)} & \textbf{Inst.}\\
  \midrule
  \S\ref{sec:bufferdesign}         & 8    & \ca{33}:\cb{33}:\cc{33}   & S, E, M & 3 & PA, RR & --      & \ca{10}:\cb{10}:\cc{10} & \ca{1}:\cb{1}:\cc{1}\\
  \S\ref{sec:simsinglereqbuff}     & 8    & \ca{33}:\cb{33}:\cc{33}   & M       & 4 & ERR    & --      & \ca{10}:\cb{10}:\cc{10} & \ca{1}:\cb{1}:\cc{1}\\
  \S\ref{sec:simreqbuffalloc}      & 4--8 & \ca{33}:\cb{33}:\cc{33}   & M       & 4 & \cstack{NRR,\\ERR, JSB} & RR & \chipstack{\ca{10/1000}:\\\cb{5/10}:\cc{1}} & \ca{4}:\cb{4}:\cc{4}\\[1pt]
  \S\ref{sec:simloadbalanceaccel}  & 8    & \ca{95}:\cb{2.5}:\cc{2.5} & S, E, M & 8 & ERR    & \cstack{RR, JSQ} & \chipstack{\ca{10/1000}:\\\cb{5/10}:\cc{1}} & \ca{3}:\cb{1}:\cc{1}\\
  \bottomrule
  \end{tabular}
\end{table}

\section{Simulator Setup}
\label{apx:simulator}
Since compiling each hardware design iteration takes hours, we explore \offrac's key design decisions with a Python simulator built on SimPy, which produced all results in Section~\ref{sec:design}. It models independent clients issuing varied requests to one multi-accelerator \offrac instance, via three components:

\noindent \textbf{Generator.} The generator simulates clients issuing requests. We abstract away network transport, assuming that fragments arrive in order but that larger requests span several fragments in distinct packets. We use the Yahoo! Cloud Serving Benchmark (YCSB)~\cite{cooper2010benchmarking} generator, configured by three properties. First, the mix of request types is set with CoreWorkload, mapping YCSB operations to accelerator requests. Second, request sizes follow one of three distributions, namely single-fragment dominant (S: 95\%, 3\%, 1\%, and 1\% of 1-, 2-, 4-, and 8-fragment requests), even mix (E: 25\% each), and multi-fragment dominant (M: 10\%, 30\%, 30\%, and 30\%). Third, 100 requests are sent closed-loop, so a client issues a new request only after receiving the previous response. Fragments arrive with a random 5--15~$\mu$s delay, and a dropped fragment is retried after a random 80--120~$\mu$s delay to avoid herd behavior. Parameters are listed in Table~\ref{tab:simulator_setup}.

\noindent \textbf{Reassembly Buffer.} This component reassembles request fragments into complete requests following the \offrac design in Section~\ref{sec:bufferdesign}, with a configurable number of buffers.

\noindent \textbf{Accelerators.} This component emulates multiple accelerator types and instances, each optionally with its own input queue of complete requests. Each is modeled by its per-fragment execution time, which scales with input size.

\section{FPGA Resources and Performance}
\label{apx:resource-tables}
Table~\ref{tab:pr_func} summarizes the accelerators used in our evaluation and partial reconfiguration analysis. The latency of A1--A3 is measured by counting clock cycles in simulation with a 4096~B input, while A4 takes a fixed input size of 24~KB.

\section{Evaluation Setup}
\label{apx:benchmarking}
We benchmark \offrac and the CPU server baseline with Libtpa~\cite{libtpa}, a DPDK-based TCP stack for high-throughput tests, via its tperf application, which we extensively modified. Clients run closed-loop tests; multiple clients maximize throughput. The server assembles each connection's requests and invokes the function once all packets of a request have arrived, mirroring \offrac. The CNN workload uses the same model for the \offrac accelerator and the software function, which runs inference with the TensorFlow C library and AVX extensions; model parameters are loaded when a connection is established and CNN is requested, so later requests on a persistent connection skip model loading.

\begin{table}[t]
  \centering
  \footnotesize
  \caption{Accelerator resource and performance summary.}
  \label{tab:pr_func}
  \setlength{\tabcolsep}{2pt}
  \resizebox{\columnwidth}{!}{%
  \begin{tabular}{@{}llcrrrrrrr@{}}
    \toprule
    \multicolumn{2}{l}{\textbf{Function}} & \textbf{Input} & \multicolumn{5}{c}{\textbf{Resources}} & \multicolumn{2}{c}{\textbf{Performance}} \\
    & & & LUT & FF & BRAM & URAM & DSP & \shortstack{Clock \\ (MHz)} & \shortstack{Latency \\ ($\mu$s)} \\
    \midrule
    \multicolumn{10}{l}{\textit{Implemented}} \\
    A1 & Top-K & 1--32KB & 9.1k  & 7.3k  & 0    & 0   & 0   & 200 & 0.41 \\
    A2 & Logit Transform & 1--32KB & 3.1k  & 7.4k  & 0    & 0   & 6   & 200 & 5.4  \\
    A3 & Min-Max Norm. & 1--32KB & 4.1k  & 6.2k  & 7.5  & 0   & 2   & 200 & 5.6  \\
    A4 & CNN   & 24KB    & 41.9k & 60.9k & 79.5 & 140 & 219 & 200 & 8600 \\
    \midrule
    \multicolumn{10}{l}{\textit{From the literature}} \\
    A5 & MM~\cite{johannes2021spclgemm} & 32KB & 60k & 88.8k & 23 & -- & 640 & 200 & 80 \\
    A6 & SVD~\cite{xilinxsvd} & 128B & 38k & 57.8k & 12 & -- & 174 & 300 & 10 \\
    A7 & PQC~\cite{beckwith2021dilithium} & 2KB & 53k & 28k & 29 & -- & 16 & 256 & 58 \\
    \bottomrule
  \end{tabular}}
\end{table}

\section{Azure Trace Processing}
\label{apx:azure-trace}
The Azure Function Invocation Trace~\cite{zhang2021faster} contains two weeks of real-world serverless invocations. As it spans many functions, we cluster invocations into accelerator classes: \textit{k}-means on invocations shorter than 1s yields three clusters, mapped to our Top-K, Logit Transform, and Min-Max Normalization accelerators; longer invocations map to CNN tasks. For request size granularity, each sub-second invocation is assigned 1, 2, 4, or 8 fragments according to its duration, using the 25th, 50th, and 75th percentiles of the corresponding cluster's duration distribution as the boundaries between the four bins; CNN invocations use 48 fragments of 512~B each, matching our CNN accelerator's fixed input size. Finally, we replay the densest hour, issuing requests at their inferred invocation times (end\_timestamp minus duration).

\section{Theoretical Analysis}
\label{apx:queue}
We estimate how long an accelerator idles when fragments are fed to it as they arrive (Immediate in Figure~\ref{fig:reassembly-sim}).
Consider a request of $k$ fragments separated by gaps $X_i\sim\Exp(\lambda)$, and a run-to-completion accelerator that spends $\tau$ on each fragment.
If the accelerator is free when fragment 1 arrives, it idles for $(X_i-\tau)^+$ after each fragment $i<k$, which by memorylessness is $e^{-\lambda\tau}/\lambda$ on average.
The total idle time is thus $\mathbb{E}[I]\le(k-1)e^{-\lambda\tau}/\lambda$, with equality for $k=2$, and the request holds the accelerator for $I+k\tau$, a share ${\eta=\mathbb{E}[I]/(\mathbb{E}[I]+k\tau)}$ of which is idle.
For $k=2$, ${\eta=e^{-\lambda\tau}/(e^{-\lambda\tau}+2\lambda\tau)}$, which is 33\% for the mean gap (35~$\mu$s) and $\tau$ (20~$\mu$s) of Figure~\ref{fig:reassembly-sim}.
Meanwhile, requests from other clients are dropped or wait behind the incomplete one.
Reassembly removes this idle time: the buffers absorb the gaps, and each request holds the accelerator only for $k\tau$.

\clearpage
\balance
\bibliographystyle{ACM-Reference-Format}
\bibliography{bibliography}

\end{document}

%% file: KAUST_Logo.tex
\definecolor{cf09021}{RGB}{240,144,33}
\definecolor{cefb61d}{RGB}{239,182,29}
\definecolor{ccccd2b}{RGB}{204,205,43}
\definecolor{c07a6aa}{RGB}{7,166,170}

\tikzset{
  kaustlogo/.pic={
  \path[fill=cf09021] (2.8502, 1.6591).. controls (2.9163, 1.7583) and (2.9494, 1.8812) .. (2.9447, 2.0041).. controls (2.9447, 2.0797) and (2.9353, 2.1554) .. (2.9116, 2.231).. controls (2.836, 2.1648) and (2.7935, 2.0703) .. (2.7935, 1.971).. controls (2.7982, 1.8623) and (2.8171, 1.7583) .. (2.8502, 1.6591);

  \path[fill=cf09021] (3.6632, 2.4673).. controls (3.6868, 2.3444) and (3.6962, 2.2215) .. (3.6962, 2.0986).. controls (3.6962, 1.6401) and (3.6064, 1.314) .. (3.4694, 1.0777).. controls (3.493, 1.073) and (3.5166, 1.073) .. (3.5403, 1.073).. controls (4.1831, 1.073) and (4.3202, 1.6685) .. (4.3202, 1.9285).. controls (4.3202, 2.3491) and (3.8948, 2.4579) .. (3.6632, 2.4673);

  \path[fill=cf09021] (3.0156, 1.9994).. controls (3.0203, 1.8481) and (2.9731, 1.7016) .. (2.8785, 1.5834).. controls (2.9022, 1.522) and (2.9353, 1.4653) .. (2.9683, 1.4085).. controls (3.096, 1.5551) and (3.1574, 1.763) .. (3.1574, 2.0325).. controls (3.1574, 2.1412) and (3.1385, 2.2499) .. (3.1054, 2.3491).. controls (3.0581, 2.3255) and (3.0156, 2.3019) .. (2.9731, 2.2688).. controls (3.0014, 2.1884) and (3.0156, 2.0939) .. (3.0156, 1.9994);

  \path[fill=cf09021] (3.2378, 2.0325).. controls (3.2378, 1.7016) and (3.148, 1.4842) .. (3.0156, 1.3424).. controls (3.0487, 1.2998) and (3.0912, 1.262) .. (3.1338, 1.2242).. controls (3.2756, 1.4038) and (3.3701, 1.6685) .. (3.3701, 2.0655).. controls (3.3701, 2.1884) and (3.3559, 2.3066) .. (3.3228, 2.4248).. controls (3.2756, 2.4106) and (3.2283, 2.3964) .. (3.1858, 2.3822).. controls (3.2189, 2.2735) and (3.2425, 2.1554) .. (3.2378, 2.0325);

  \path[fill=cf09021] (3.4694, 2.0655).. controls (3.4694, 1.6401) and (3.3654, 1.3566) .. (3.2141, 1.1628).. controls (3.2567, 1.1344) and (3.3039, 1.1155) .. (3.3512, 1.1013).. controls (3.4883, 1.3235) and (3.5781, 1.6449) .. (3.5781, 2.0986).. controls (3.5781, 2.2215) and (3.5639, 2.3444) .. (3.5403, 2.4673).. controls (3.5025, 2.4626) and (3.4599, 2.4579) .. (3.4126, 2.4484).. controls (3.4505, 2.3255) and (3.4694, 2.1979) .. (3.4694, 2.0655);

  \path[fill=cefb61d] (2.1837, 3.0298).. controls (2.5382, 3.0298) and (2.8313, 2.8927) .. (3.0203, 2.6706).. controls (3.0581, 2.7226) and (3.0912, 2.7793) .. (3.1196, 2.836).. controls (2.9494, 3.0487) and (2.6895, 3.1905) .. (2.335, 3.1905).. controls (2.0466, 3.181) and (1.763, 3.129) .. (1.4889, 3.0298).. controls (1.4936, 2.992) and (1.5078, 2.9542) .. (1.5267, 2.9211).. controls (1.7394, 2.992) and (1.9616, 3.0298) .. (2.1837, 3.0298);

  \path[fill=cefb61d] (3.2141, 3.3039).. controls (3.1905, 3.6253) and (3.0203, 4.1547) .. (2.4295, 4.1547).. controls (1.971, 4.1547) and (1.6969, 3.7246) .. (1.5692, 3.3937).. controls (1.867, 3.4883) and (2.1743, 3.5403) .. (2.4815, 3.545).. controls (2.7462, 3.5544) and (3.0062, 3.4694) .. (3.2141, 3.3039);

  \path[fill=cefb61d] (1.6732, 2.7367).. controls (1.7914, 2.7651) and (1.9143, 2.7745) .. (2.0372, 2.7745).. controls (2.2357, 2.7745) and (2.6186, 2.7367) .. (2.8407, 2.4957).. controls (2.888, 2.5288) and (2.9305, 2.5618) .. (2.9683, 2.6044).. controls (2.7982, 2.8171) and (2.5193, 2.9494) .. (2.1837, 2.9494).. controls (1.9757, 2.9494) and (1.7678, 2.9163) .. (1.5645, 2.8502).. controls (1.5976, 2.8076) and (1.6354, 2.7698) .. (1.6732, 2.7367);

  \path[fill=cefb61d] (2.7793, 2.4626).. controls (2.5713, 2.6753) and (2.2215, 2.7084) .. (2.0419, 2.7084).. controls (1.9427, 2.7084) and (1.8434, 2.6989) .. (1.7489, 2.6847).. controls (1.9757, 2.5193) and (2.3019, 2.4011) .. (2.5146, 2.4011).. controls (2.5997, 2.3964) and (2.6942, 2.4153) .. (2.7793, 2.4626);

  \path[fill=cefb61d] (2.335, 3.2898).. controls (2.7036, 3.2898) and (2.9778, 3.148) .. (3.1621, 2.9353).. controls (3.1858, 3.0062) and (3.2047, 3.0771) .. (3.2141, 3.1527).. controls (3.0534, 3.3181) and (2.8171, 3.4316) .. (2.4815, 3.4316).. controls (2.1554, 3.4221) and (1.8292, 3.3607) .. (1.522, 3.2567).. controls (1.5125, 3.2189) and (1.5031, 3.1763) .. (1.4936, 3.1385).. controls (1.763, 3.233) and (2.0466, 3.285) .. (2.335, 3.2898);

  \path[fill=ccccd2b] (0.7894, 2.3964).. controls (0.7894, 2.6658) and (0.9595, 2.8596) .. (1.1958, 3.0014).. controls (1.1628, 3.0251) and (1.1249, 3.044) .. (1.0871, 3.0534).. controls (0.865, 2.9116) and (0.7137, 2.7178) .. (0.7137, 2.472).. controls (0.7137, 2.2499) and (0.7941, 1.9663) .. (0.9406, 1.6921).. controls (0.9642, 1.6921) and (0.9879, 1.6969) .. (1.0115, 1.7063).. controls (0.8697, 1.9474) and (0.7894, 2.2026) .. (0.7894, 2.3964);

  \path[fill=ccccd2b] (0.9784, 2.3208).. controls (0.9784, 2.5524) and (1.1297, 2.7178) .. (1.3329, 2.8313).. controls (1.3187, 2.8644) and (1.2951, 2.8974) .. (1.2715, 2.9258).. controls (1.0493, 2.8029) and (0.8886, 2.6327) .. (0.8886, 2.3917).. controls (0.8886, 2.2073) and (0.9642, 1.971) .. (1.0966, 1.7394).. controls (1.1202, 1.7583) and (1.1391, 1.7725) .. (1.158, 1.7961).. controls (1.0446, 1.9852) and (0.9831, 2.1743) .. (0.9784, 2.3208);

  \path[fill=ccccd2b] (0.9312, 3.0865).. controls (0.9028, 3.0912) and (0.8744, 3.0912) .. (0.8461, 3.0912).. controls (0.5767, 3.0912) and (0.0, 2.8691) .. (0.0, 2.4579).. controls (0.0, 2.0183) and (0.4868, 1.7489) .. (0.8035, 1.6969).. controls (0.6665, 1.9757) and (0.5908, 2.2546) .. (0.5908, 2.472).. controls (0.5956, 2.7273) and (0.7279, 2.9305) .. (0.9312, 3.0865);

  \path[fill=ccccd2b] (1.1628, 2.2499).. controls (1.1628, 2.4011) and (1.2431, 2.5335) .. (1.3849, 2.6233).. controls (1.3849, 2.6706) and (1.3755, 2.7178) .. (1.3613, 2.7604).. controls (1.1817, 2.6611) and (1.0588, 2.5193) .. (1.0588, 2.3255).. controls (1.0682, 2.1648) and (1.1202, 2.0088) .. (1.2053, 1.867).. controls (1.2242, 1.8954) and (1.2384, 1.9285) .. (1.2526, 1.9568).. controls (1.2006, 2.0466) and (1.1675, 2.1459) .. (1.1628, 2.2499);

  \path[fill=ccccd2b] (1.2337, 2.2499).. controls (1.2384, 2.1743) and (1.2573, 2.1034) .. (1.2904, 2.0372).. controls (1.3471, 2.1979) and (1.3802, 2.3681) .. (1.3896, 2.5382).. controls (1.2904, 2.472) and (1.2337, 2.3633) .. (1.2337, 2.2499);

  \path[fill=c07a6aa] (2.2593, 1.366).. controls (1.9663, 1.3707) and (1.6874, 1.4842) .. (1.4794, 1.6827).. controls (1.4416, 1.6685) and (1.4085, 1.6543) .. (1.3755, 1.6354).. controls (1.5976, 1.3755) and (1.9237, 1.1628) .. (2.3161, 1.1628).. controls (2.4815, 1.158) and (2.6422, 1.1911) .. (2.7887, 1.262).. controls (2.7651, 1.3282) and (2.732, 1.3896) .. (2.6895, 1.4416).. controls (2.5477, 1.3896) and (2.4059, 1.3613) .. (2.2593, 1.366);

  \path[fill=c07a6aa] (1.0021, 1.3755).. controls (0.865, 1.2384) and (0.7799, 1.0871) .. (0.7799, 0.9312).. controls (0.7799, 0.3356) and (1.7063, 0.0) .. (2.0655, 0.0).. controls (2.4909, 0.0) and (2.6895, 0.2694) .. (2.7745, 0.553).. controls (2.6564, 0.5247) and (2.5335, 0.5105) .. (2.4106, 0.5105).. controls (1.815, 0.5152) and (1.3235, 0.9075) .. (1.0021, 1.3755);

  \path[fill=c07a6aa] (1.919, 1.7772).. controls (1.7914, 1.7772) and (1.6685, 1.7536) .. (1.5503, 1.711).. controls (1.7441, 1.5362) and (1.9994, 1.4369) .. (2.2593, 1.4369).. controls (2.387, 1.4322) and (2.5146, 1.4558) .. (2.6327, 1.4983).. controls (2.4059, 1.73) and (2.0514, 1.7772) .. (1.919, 1.7772);

  \path[fill=c07a6aa] (2.3113, 1.0824).. controls (1.8907, 1.0824) and (1.5362, 1.3187) .. (1.2998, 1.5976).. controls (1.2715, 1.5834) and (1.2478, 1.5692) .. (1.2195, 1.5503).. controls (1.4794, 1.1958) and (1.8812, 0.8981) .. (2.3586, 0.8981).. controls (2.5193, 0.8933) and (2.68, 0.9264) .. (2.8265, 0.9879).. controls (2.8265, 1.054) and (2.8171, 1.1202) .. (2.8029, 1.1817).. controls (2.6469, 1.1155) and (2.4815, 1.0777) .. (2.3113, 1.0824);

  \path[fill=c07a6aa] (2.3586, 0.7988).. controls (1.8481, 0.7988) and (1.418, 1.1202) .. (1.1391, 1.4983).. controls (1.1202, 1.4842) and (1.1013, 1.4653) .. (1.0824, 1.4511).. controls (1.3849, 1.0068) and (1.8481, 0.6334) .. (2.4106, 0.6334).. controls (2.5429, 0.6334) and (2.6753, 0.6523) .. (2.8029, 0.6901).. controls (2.8171, 0.7563) and (2.8218, 0.8224) .. (2.8265, 0.8886).. controls (2.68, 0.8272) and (2.5193, 0.7988) .. (2.3586, 0.7988);
}
}

\newcommand\kausticon{\mbox{\scalerel*{
\begin{tikzpicture}[yscale=1,transform shape]
\pic{kaustlogo};
\end{tikzpicture}
}{|}}}

%% file: comp-table.tex
% ========== Original compact table (commented out) ==========
\begin{table}[t]
    \caption{Comparison of \offrac with existing work.
        \normalfont{\emph{TP}: hardware transport-protocol support (\emph{Partial}: NIC offloads or a custom/RPC transport, not a complete in-hardware stack).
        \emph{RG}: request-level granularity (\emph{Specific}: application-specific messages, not a general request abstraction).
        \emph{MT}: multi-tenancy.
        \emph{AO}: accelerator offload (\emph{Packet}: packet-level network functions only; \emph{Static}/\emph{Dynamic}: application accelerators without/with runtime reconfiguration).
        \emph{HD}: hostless deployment.}}
    \label{tab:comparison}
    \centering
    \footnotesize
    \renewcommand{\arraystretch}{0.9}
    \begin{tabular}{@{}lccccc@{}}
    \toprule
    \textbf{} & \textbf{TP} & \textbf{RG} & \textbf{MT} & \textbf{AO} & \textbf{HD} \\
    \midrule
    RingLeader~\cite{lin2023ringleader}    & \xmark  & \xmark  & \xmark & \xmark  & \xmark  \\
    PANIC~\cite{lin2020panic}      & \xmark  & \xmark  & \cmark     & Packet & \xmark  \\
    SuperNIC~\cite{lin2024supernic}     & \xmark  & \xmark  & \cmark     & Packet & \xmark  \\
    ClickNP~\cite{Li2016ClickNP}        & \xmark & \xmark  &  \xmark    & Packet  & \xmark  \\
    Beehive~\cite{lim2024beehive}        & \cmark & \xmark  & \xmark      & Packet & \cmark  \\
    TOEs~\cite{mogul2003tcp} & \cmark & \xmark & \xmark & \xmark & \xmark \\
    Pigasus~\cite{zhao2020pigasus} & \cmark & \xmark & \cmark & Static & \xmark \\
    HFT~\cite{leber2011hft} & \cmark & Specific & \xmark & Static & \xmark \\
    BlastFunction~\cite{bacis2020blastfunction} & \xmark  & \xmark  & \cmark     & Static     & \xmark  \\
    Molecule~\cite{Dong2022Serverless} & \xmark & \xmark & \cmark & Static & \xmark\\
    F3~\cite{mainas_f3_2025} & \xmark & \xmark & \cmark & Dynamic & \xmark\\
    Harmonia~\cite{Li2025Harmonia}& \xmark & \xmark & \cmark & Static & \xmark\\
    Ens{\=o}~\cite{sadok2023enso}        & \cmark & \xmark & \cmark     & \xmark      & \xmark  \\
    nanoPU~\cite{Stephen2021NanoPU} &\cmark & \cmark & \cmark & \xmark & \xmark \\
    RpcNIC~\cite{zhang_rpcnic_nodate} & Partial & \cmark & \cmark & Dynamic & \xmark \\
    Split gRPC~\cite{ramos2024splitgrpc} & \xmark & \cmark & \cmark & \xmark & \xmark \\
    FlexDriver~\cite{FlexDriver2022Eran} & \xmark &\cmark &\cmark &Static &\xmark\\
    NEBULA~\cite{sutherland_nebula_2020} &Partial & \cmark & \cmark & \xmark & \xmark \\
    Arca~\cite{srivatsan2026arca} & \xmark & \cmark & \cmark & \xmark & \xmark \\
    Rain~\cite{ma2026rain} & \cmark & \cmark & \cmark & \xmark & \xmark \\
    Coyote v1 \& v2~\cite{korolija2020OS,Ramhorst2025Coyote}     & \cmark & \xmark  & \cmark     & Dynamic     & \xmark  \\ 
    Strega~\cite{maschi2024strega} & \cmark & \cmark & \cmark & Static & \xmark\\\midrule
    \textbf{\offrac}   & \textbf{\cmark} & \textbf{\cmark} & \textbf{\cmark}     & \textbf{Dynamic}     & \textbf{\cmark} \\
    \bottomrule
    \end{tabular}
\end{table}